\documentclass[reprint,
               amsmath,
               amssymb,
               aps,
               footinbib]{revtex4-2}
\usepackage{graphicx}
\usepackage{dcolumn}
\usepackage{bm}
\usepackage{xcolor}
\usepackage{caption}
\usepackage{subcaption}
\usepackage[colorlinks, citecolor=blue]{hyperref}
\usepackage{mathtools}
\makeatletter
\newcommand{\revtextableofcontents}{
  \begingroup
    \addtocontents{toc}{\protect\setcounter{tocdepth}{-1}}
    \par{\large\bfseries\tocname}\par\vspace{0.5ex}
    \@starttoc{toc}
    \addtocontents{toc}{\protect\setcounter{tocdepth}{2}}
  \endgroup
}
\makeatother

\usepackage{footnote}
\makesavenoteenv{figure*}

\usepackage{color,soul}

\newcounter{para}

\AtBeginDocument{\def\selectlanguage#1{}}

\usepackage{microtype}
\DisableLigatures[f]{encoding = *, family = * }
\usepackage[utf8]{inputenc}
\usepackage{graphicx}
\usepackage{adjustbox}
\usepackage{dcolumn}
\usepackage{bm}
\usepackage{ulem}
\usepackage{amsmath}
\usepackage{amssymb}
\usepackage{mathtools}
\usepackage{tabularx}
\usepackage{wrapfig}
\usepackage{nameref}
\usepackage{siunitx}
\usepackage{ulem}
\usepackage{tocloft}

\usepackage{setspace}
\usepackage{multirow,array}
\usepackage{graphicx}
\usepackage{amsmath}
\usepackage{amssymb}
\usepackage{float}
\usepackage{tikz}
\usepackage{xcolor}
\usepackage{caption}
\usepackage[most]{tcolorbox}
\usepackage{braket}
\usepackage{bbm}
\usepackage{subcaption}

\begin{document}

\title{Thermal fluctuations set fundamental limits on ion channel function}

\author{Jose M. Betancourt}
 \email{jose.betancourtvalencia@yale.edu}
\affiliation{Department of Physics, Yale University}
\affiliation{Quantitative Biology Institute, Yale University}

\author{Benjamin B. Machta}
 \email{benjamin.machta@yale.edu}
\affiliation{Department of Physics, Yale University}
\affiliation{Quantitative Biology Institute, Yale University}

\date{April 4, 2026}

\begin{abstract}
Voltage-gated ion channels are essential for propagating signals in neurons. Each channel senses the local membrane potential created by nearby ions. Fluctuations in these ions introduce two fundamental noise sources: (i) shot noise, from the discreteness of ionic charge, and (ii) Johnson–Nyquist noise, from long‑wavelength thermal fluctuations of the electric field. We show that, for an individual channel, shot noise dominates and sets an intrinsic limit to voltage sensing. On the $10$ $\mu$s timescales relevant to channel gating, this limit corresponds to an accuracy of about $10$ mV -- close to measured channel sensitivities. When signals from many channels are aggregated, Johnson–Nyquist noise eventually overtakes shot noise and bounds the total information that can be sensed from the environment. This transition occurs at an ion channel density of $< 1$ channel/$\mu$m$^2$ for slow signals and around $10^2-10^4$ channels/$\mu$m$^2$ for signals with $10$ $\mu$s timescales, both of which are within the range of experimentally-measured densities for somas and axon initial segments, respectively. These results provide design principles for single‑channel architecture and collective sensing and suggest that neuronal computation is ultimately constrained by thermal fluctuations.
\end{abstract}

\maketitle

\addtocontents{toc}{\protect\setcounter{tocdepth}{-1}}

\section{Introduction}
Biological function is shaped by constraints, sometimes arising from simple physical considerations.  In classic work, Berg and Purcell showed that thermal fluctuations impose an absolute bound on how accurately a cell can measure concentrations \cite{berg_physics_1977}.  Their bound takes as input only the diffusion constant of chemical species and the geometry of the sensor.  This initial work has inspired many followups on the limits of chemical sensing \cite{bialek_physical_2005, kaizu_berg-purcell_2014, hu_physical_2010, endres_accuracy_2008, andrews_optimal_2006}.  However, different biological sensors are subject to thermal fluctuations with qualitatively different structure~\cite{bryant_physical_2023}. Here we consider the physical limitations on sensing electrical signals.

There are important qualitative differences between ligand-based communication and electrical communication. Unlike ligands undergoing diffusion, ions strongly interact with each other through long-range Coulomb interactions.  These long range forces allow electrical signals to propagate far faster than individual components diffuse.  They also lead to a more complex structure of thermal fluctuations.  At long lengths thermal fluctuations obey emergent dynamics of a continuum model, known as Johnson-Nyquist noise. At these large scales, fluctuations from the diffusion of charge carriers are effectively screened and strongly suppressed. In contrast, at short length scales these fluctuations manifest as shot noise. The properties of the fluctuations of charge in finite volumes has been studied extensively, both for static \cite{martin_charge_1980, bekiranov_fluctuations_1998, levesque_no_2000, jancovici_no_2003, kim_charge_2008} and dynamic \cite{caillol_theoretical_1986, caillol_dielectric_1987, zorkot_power_2016, lesnicki_molecular_2021, hoang_ngoc_minh_ionic_2023} phenomena. Because electrical communication is mediated by relatively small sensors separated by large distances, the behavior of fluctuations in both the short and long length regimes will turn out to be relevant.

In previous work, we \cite{bryant_physical_2023} analyzed the role of noise structure in designing communication channels for processes at different length and time scales. In particular, we studied how electrical communication can be used to send signals quickly over long distances, but at high energetic costs. We used a continuum model of electrical dynamics, allowing us to characterize signal  propagation between channels.  However, this continuum approximation only captured the Johnson-Nyquist contribution to noise.  

By starting with a more microscopic model, here we study the fundamental limits that shot noise and Johnson-Nyquist noise place on electrical sensing in the cellular environment.  We first consider a model for a single ion channel, where we find that shot noise dominates, and provides a fundamental bound on the sensitivity of any voltage gated channel.  Ion channels involved in action potential propagation operate within an order of magnitude of our bound, suggesting this physical constraint is binding.

Following Berg and Purcell's analysis of chemical sensing, we next study how noise limits an idealized measurement performed by multiple ion channels. We find that Johnson-Nyquist noise dominates for large numbers of channels, saturating at an accuracy set by geometry and the coarse-grained electrical properties of the cell. Analogous to Berg and Purcell, we show that this information can be recovered by sparsely populating the cell surface with ion channels.

\section{Dynamics of charged ions in the presence of a membrane}

To model the dynamics of ions in a solution, we use the Poisson-Nernst-Planck (PNP) model~\cite{macdonald_binary_1974, kornyshev_conductivity_1981, bazant_diffuse-charge_2004, row_spatiotemporal_2025, farhadi_capacitive_2025}. In this model, each ion experiences two kinds of forces: electrical interactions with the other ions and entropic forces. The resulting currents ${\bf j}_i$ for the density, $n_i$ of individual ionic species is given by
\begin{align}
    {\bf j}_i = \frac{z_ieDn_i}{k_BT } {\bf E} - D \nabla n_i,
\end{align}
where ${\bf E}$ is the electric field, $z_i$ is the valence of species $i$, $e$ is the elementary charge, and $D$ is the diffusion constant of ions, here assumed equal.  The total electric current obeys
\begin{align}
    {\bf J} = \alpha {\bf E} - D \nabla \rho,
\end{align}
where $\rho$ is the electric charge density and $\alpha=\frac{D}{k_BT}\sum_in_i(ez_i)^2$ is the conductivity of the medium. To close the dynamics, we use Gauss's law: $\nabla \cdot {\bf E} = \rho/\epsilon$, where $\epsilon$ is the permittivity of the bulk. With this, the charge dynamics can be expressed as
\begin{align} \label{eq:pnp_charge_dynamics}
    \tau_D \partial_t \rho = - \rho + l_D^2 \nabla^2 \rho,
\end{align}
where $\tau_D \coloneqq \epsilon/\alpha$ and $l_D^2 \coloneqq D \epsilon/\alpha$ are the Debye time and the Debye length, respectively. In water, these are $\tau_D \approx 1$ ns and $l_D \approx 0.7$ nm. Equation \eqref{eq:pnp_charge_dynamics} implies that long-wavelength inhomogeneities decay with a characteristic time scale of $\tau_D$. Therefore, for slow time scales the effective bulk charge is $\rho \approx 0$, but charge currents satisfying ${\bf J} \approx \alpha {\bf E}$ can still exist. This is the basis for a continuum treatment of electrical dynamics.

To study communication mediated by electric charges we introduce a membrane into our system, which allows charge inhomogeneities to persist. We model the membrane as an infinite impenetrable slab of thickness $d$ with permittivity $\epsilon_m$. As we show in the Supplemental Material, the dynamics of charge in the presence of a membrane can be described by the 2D effective model studied in \cite{bryant_physical_2023}. To obtain these dynamics, consider a model in which the membrane is a 2D capacitor with capacitance per unit area $c$. The charge on the capacitor can be described by the 2D charge density field $\lambda({\bf x})$. By considering the currents generated by the electric field, one obtains the following non-local dynamics for $\lambda$:
\begin{align}
    \partial_t \lambda({\bf k}, t) = - \frac{\alpha k}{2 c} \lambda({\bf k}, t).
\end{align}
In our original model, the 3D charge density is effectively $\lambda$ spread out over a Debye length: $\rho({\bf x}, z) \approx l_D^{-1} \lambda({\bf x}) \text{sgn}(z) e^{-|z|/l_D}$. The capacitance in our 2D model is the effective capacitance of the membrane and two Debye layers in parallel: $c^{-1} = c_0^{-1} + 2l_D/\epsilon$, where $c_0 \coloneqq \epsilon_m/d$. With these effective dynamics, the effective decay time of the mode ${\bf k}$ is $2c/\alpha k$, which can be much larger than $\tau_D$ for long wavelengths. Thus, the localization of charge in the Debye layer can be used for stable communication over long distances.

\section{Physical limits on voltage sensing} \label{sec:noise}
\begin{figure}
    \centering
    \includegraphics[width = 0.5 \linewidth]{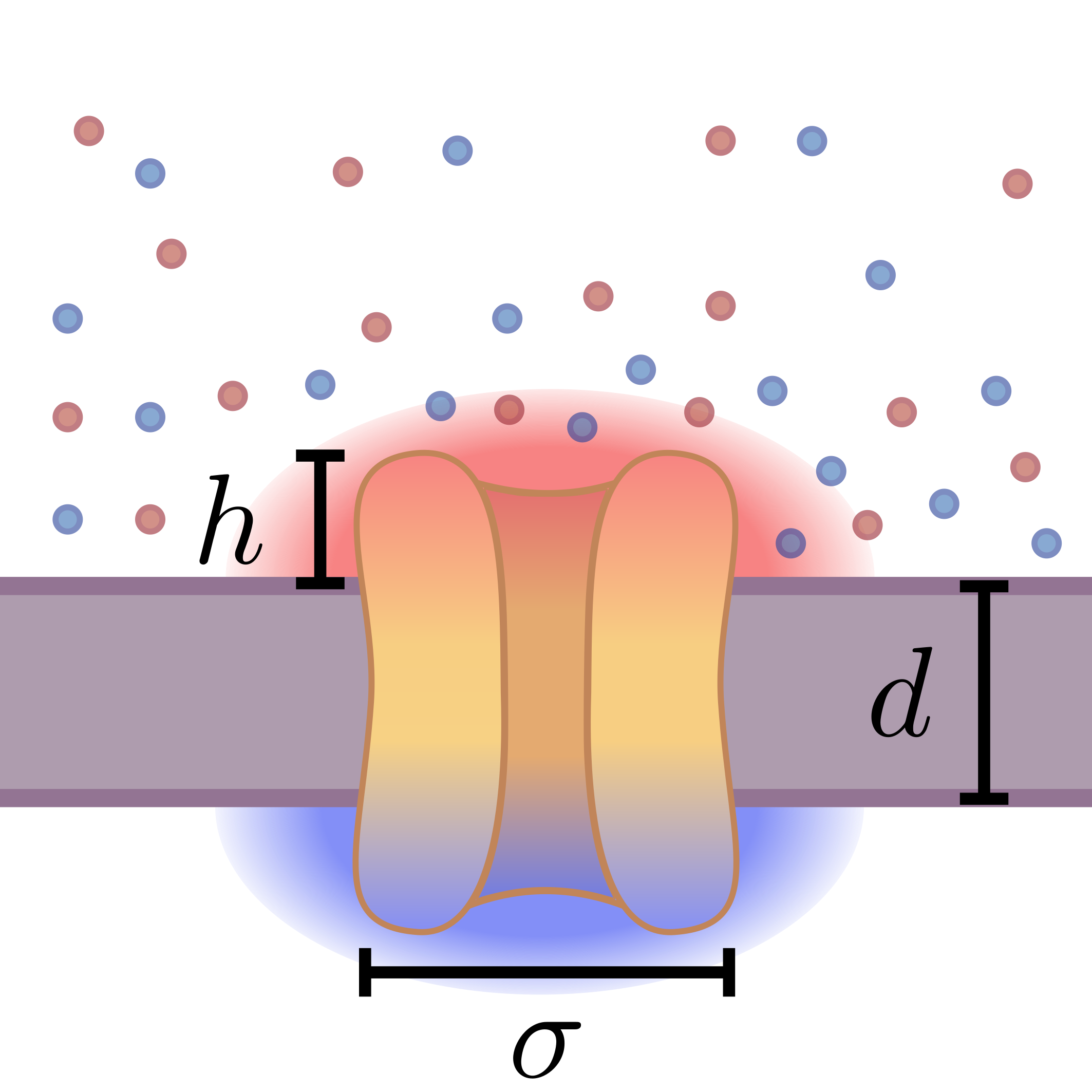}
    \caption{Charge sensed by ion channels is generated by a discrete distribution of ions. The ion channel senses charge in a region of width $\sigma$ and height $h$. It is embedded in a membrane of thickness $d$.}
    \label{fig:ion_channel}
\end{figure}

We have shown that the presence of a membrane allows signals to be sent over long distances. In neurons, these signals are sensed by ion channels, which detect and amplify voltage changes to generate action potentials \cite{hodgkin_quantitative_1952} and other dynamics. We model an ion channel as a sensor that can measure charge in a small volume near the membrane. The channel makes a time-dependent measurement of the charge in its vicinity:
\begin{align}
    M(t) = \int d^3{\bf r} \, e^{-\lVert{\bf x} \rVert^2/2 \sigma^2} e^{-|z|/h} \text{sgn}(z) \rho({\bf r}, t),
\end{align}
where $\sigma$ is the width of the channel and $h$ is its height (see Figure \ref{fig:ion_channel}).  We note some ambiguity in our choice of observable $M(t)$, which is not present in the analogous calculation for the noise from particle diffusion first considered by Berg and Purcell.  In particle diffusion, both signal and noise are carried by the same bulk concentration modes.  In our electrical dynamics only the surface modes carry signal, while both bulk and surface modes contribute to noise.  Here we optimize over sensor height $h$ to minimize the relative strength of these bulk modes, but we cannot rule out that a qualitatively different measurement might be even less sensitive, see discussion and supplement.

\begin{figure*}
    \centering
    \includegraphics[width=\linewidth]{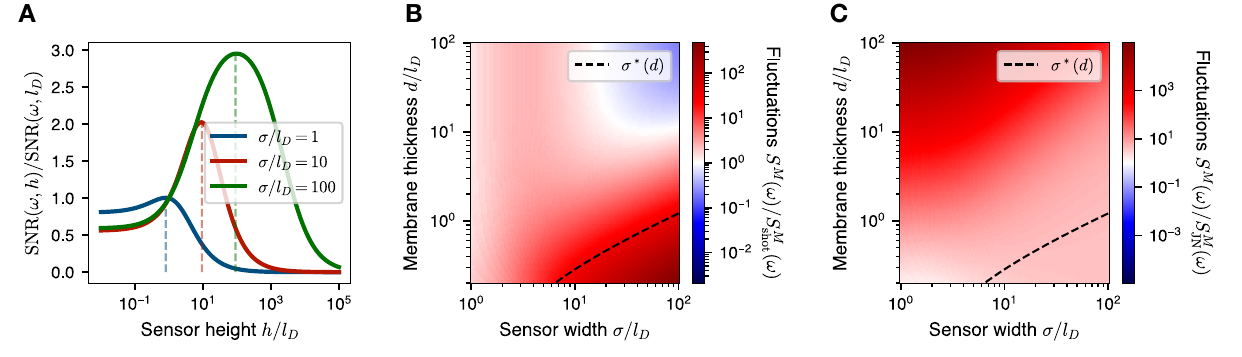}
    \caption{Numerical analysis of fluctuations for the parameters in Table \ref{tab:parameters}. (A) Scaling of signal-to-noise ratio with sensor height $h$. Dashed lines represent the unique optimal heights. (B) Comparison of fluctuations to shot noise. Values are shown at the optimal height $h^*(\sigma, d)$. The dashed line represents the analytical prediction of where the two contributions are equal. (C) Comparison of fluctuations to Johnson-Nyquist noise.}
    \label{fig:noise_numerics}
\end{figure*}

We assume that the fluctuations in $M$ are equilibrium thermal fluctuations from the movement of the ions. To obtain them, we use the fluctuation-dissipation theorem, which relates the response of $M$ to a perturbation in the electric potential to its equilibrium fluctuations. Specifically, we perturb the free energy of the system by introducing a fictitious potential that has the same shape as the channel. We then derive the induced dynamics of the charge to obtain the response of the measurement to the perturbation (see Supplemental Material for details). While the expression for the fluctuations is quite complicated, we can make progress by analyzing its scaling with channel width and identifying the leading sources of noise.

The height of the channel $h$ determines how far into the bulk it extends. Since charge inhomogeneities are localized within $l_D$ of the membrane, we would expect the height of the channel to be of order $l_D$ to capture any signals sent through this layer. In the Supplemental Material we characterize the signal-to-noise ratio (SNR) of a communication channel where a signal is sent from a large distance and is sensed by our sensor. This SNR is shown as a function of height in Figure \ref{fig:noise_numerics}.A. While the optimal height increases with $\sigma$, the gain from having a channel larger than a few Debye lengths is not very large. This is because the signal quickly saturates for heights larger than $l_D$, so a good performance can be achieved by having $h$ be of the order of $l_D$.

To gain intuition for what determines the magnitude of fluctuations, we study a sensor of height $l_D$. Since ion channels are usually larger than the Debye length ($\sigma > l_D$), we study the leading-oder contributions when we expand the noise in $l_D/\sigma$. Keeping the two leading-order terms, we identify two contributions to the measurement fluctuations:
\begin{align}
    S^M(\omega) \approx S^M_{\text{JN}}(\omega) + S^M_{\text{shot}}(\omega) \quad (\omega \ll \tau_D^{-1}),
\end{align}
where
\begin{align}
    S^M_{\text{JN}}(\omega) &\coloneqq 4 \pi^{3/2} \frac{\sigma^3 c^2}{\alpha} k_B T, \nonumber \\
    S^M_{\text{shot}}(\omega) &\coloneqq \frac{\pi}{4} \frac{\sigma^2 \epsilon \tau_D}{l_D} k_B T.
\end{align}
We identify $S^M_{\text{JN}}(\omega)$ with Johnson-Nyquist noise, since it coincides with the fluctuations obtained in \cite{bryant_physical_2023} for a continuum model of charge dynamics. Note that this contribution depends on the effective capacitance $c$. When $c_0 \ll \epsilon/l_D$ (as is the case for our parameters), this term is heavily suppressed, so the noise in the continuum model becomes small. The second source of noise captures the contribution of shot noise to fluctuations. This term does not depend on the properties of the membrane, and dominates Johnson-Nyquist noise when the membrane capacitance is small. To verify that these sources of noise capture the behavior of fluctuations, we performed a parameter sweep on membrane height $d$ and sensor width $\sigma$, shown in Figures \ref{fig:noise_numerics}.B and \ref{fig:noise_numerics}.C. For the explored parameter range, our predictions for the total noise are within an order of magnitude of the numerical solution. This means that Johnson-Nyquist and shot noise capture the relevant structure of the fluctuations that affect our sensor.

\begin{table}
    \centering
    \begin{tabular}{c|cc} \hline \hline
        Quantity & Symbol & Value \\ \hline
        Conductivity & $\alpha$ & $10^{-6}$ S/$\mu$m \cite{wang_role_2013} \\
        Permittivity & $\epsilon$ & $80$ $\epsilon_0$ \\
        Membrane thickness & $d$ & 5 nm \cite{eyal_unique_2016} \\ 
        Capacitance/area & $c$ & $1$ $\mu$F/cm$^2$ \cite{gentet_direct_2000} \\
        Ion channel width & $\sigma$ & 1-5 nm \\
        Radius of soma & $R$ & 10 $\mu$m \\
        Membrane leak time & $\tau_{\text{leak}}$ & 1 ms \\ \hline \hline
    \end{tabular}
    \caption{Parameters used in examples. The range in channel sizes is meant to capture the possibility of having the entire channel or specific sub-components acting as our sensor.}
    \label{tab:parameters}
\end{table}

To gain some intuition on the scale of Johnson-Nyquist and shot noise, we evaluate them for some physiologically-relevant parameters. These are shown in Table \ref{tab:parameters}. First, note that $c \approx 10^{-2} \epsilon/l_D$, so we are deep in the regime where the effective capacitance is dominated by the membrane capacitance $c_0$. For these parameters, we have $S^M_{\text{shot}}(\omega)/S^M_{\text{JN}}(\omega) \approx 50 - 250$, meaning that fluctuations are dominated by shot noise. 

The structure of fluctuations determines the sensing accuracy of a sensor. In the case of ion channels, we can characterize how accurately they can detect voltage. In a continuum model, the voltage is related to the 2D charge density by $V({\bf x}) = \lambda({\bf x})/c$. By relating our measurement of charge to this voltage field, we can characterize the error in the ion channel's estimate of the voltage. If the measurement is taken over a time $t_{\text{meas}} \gg \tau_D$, the error in the measurement is
\begin{align}
    \delta V \approx V_0 \left( \frac{t_{\text{meas}}}{\tau_D} \right)^{-1/2}, \quad V_0 \coloneqq \left( \frac{1}{16 \pi} \frac{\epsilon k_B T}{c^2 \sigma^2 l_D} \right)^{1/2}.
\end{align}
We can interpret $V_0$ as the resolution with which a measurement can be made within a Debye time. For the parameters in Table \ref{tab:parameters}, this resolution is $V_0 \approx 180-900$ mV. During action potentials, ion channels respond to changes of the order of 10 mV. This resolution can be achieved with a measurement time of $t_{\text{meas}} \approx 0.1-10$ $\mu$s. Single-channel measurements have shown that ion channels react to changes in $\sim 10$ mV in $\sim 100$ $\mu$s \cite{aldrich_reinterpretation_1983}. Additionally, ion channels must react to changes in voltage in timescales of order $10$ $\mu$s in fast action potential propagation, as in the nodes of Ranvier. Thus, our shot noise bound sets a limit to sensing that is within an order of magnitude of observed ion channel response times.

\section{The perfect instrument}
\begin{figure}
    \centering
    \includegraphics[width = 0.7 \linewidth]{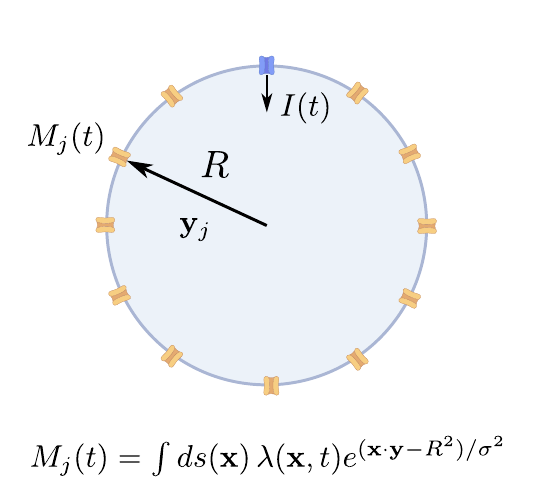}
    \caption{Communication scheme in the soma. A time-dependent current is injected from a source. This signal propagates through the charge density field $\lambda$ and is sensed by $N$ ion channels distributed throughout the sphere. In the spherical integrals, vectors on the sphere correspond to ${\bf x} \coloneqq (R \sin(\theta) \cos(\phi), R \sin(\theta) \sin(\phi), R \cos(\theta))$ and the area element is $ds({\bf x}) \coloneqq R^2 \sin(\theta) \, d\theta \, d\phi$.}
    \label{fig:spherical_soma}
\end{figure}

Having characterized the noise structure of measurements by ion channels, we can ask how integrating information over many ion channels could help overcome shot noise. To do this, we study a neuron that uses spatially separated sensors to sense some signal (Figure \ref{fig:spherical_soma}). We begin by characterizing the ``perfect instrument'': a cell that can measure charge over its entire surface. Such an instrument would be immune to shot noise, since it can average out many measurements spatially. We show, however, that it is still subject to Johnson-Nyquist noise.

We model the soma of a neuron as a spherical membrane of radius $R$ with capacitance per unit area $c$. As before, it is embedded in a bulk medium with conductivity $\alpha$. The charge density on the membrane is $\lambda({\bf x})$, where ${\bf x}$ now represent coordinates on the surface of the sphere. We assume there are charge leaks across the surface, characterized by the leak current $J_{\textrm{leak}}({\bf x},t) = - \tau_{\text{leak}}^{-1} \lambda({\bf x},t)$. This current sets the characteristic time scale of the charge dynamics.

We consider a signal generated by a ``sender'' (which could be an ion channel, a dendrite, or any other source of external information) located at the north pole of the sphere. The sender generates an input current $I(t)$ by transporting ions through the membrane, which results in an input current $J({\bf x}, t)$. The dynamics of the charge density can be easily written in terms of spherical harmonics:
\begin{align}
    \partial_t \lambda_\ell^m(t) = - \frac{1}{\tau_l} \lambda_\ell^m(t) + J_\ell^m(t),
\end{align}
where $\tau_\ell$ is the RC time of each harmonic components. For the zeroth harmonic, this is $\tau_0 = \tau_{\text{leak}}$. For higher harmonics, the RC times scale as $\tau_\ell \approx \frac{Rc}{\alpha} \frac{2\ell + 1}{\ell (\ell+1)}$. For the parameters in Table \ref{tab:parameters}, we have $\tau_1 \approx 10^{-4} \tau_0$, so all other harmonics decay much faster than the zeroth one.

A cell that can measure charge on its entire surface can make the measurement $M_{\text{cell}}(t) = \int ds({\bf x}) \, \lambda({\bf x}, t)$. Since the charge dynamics are linear, this measurement has a linear response behavior to the input signal, given by $\langle M_{\text{cell}}(\omega) \rangle = \chi^{MI}_{\text{cell}}(\omega) I(\omega)$. The response kernel $\chi^{MI}_{\text{cell}}(\omega)$ characterizes the strength of the signal obtained by making the measurement $M_{\text{cell}}$. The quality of our communication channel can be quantified by the signal-to-noise ratio $\text{SNR}_{\text{cell}}(\omega) = |\chi^{MI}_{\text{cell}}(\omega)|^2 S^I(\omega)/S^M_{\text{cell}}(\omega)$, where $S^I(\omega)$ is the input power spectrum and $S^M_{\text{cell}}(\omega)$ is the spectrum of fluctuations. Note that since we are using a continuum model of charge dynamics, the noise we obtain corresponds to Johnson-Nyquist noise. For this whole-cell measurement, the response kernel, fluctuation spectrum and signal-to-noise ratio are
\begin{gather}
    \chi^M_{\text{cell}}(\omega) = \frac{\tau_{\text{leak}}}{1 + i \omega \tau_{\text{leak}}}, \quad S^M_{\text{cell}}(\omega) = \frac{8 \pi R^2 c \tau_{\text{leak}}}{(1 + \omega^2 \tau_{\text{leak}}^2)}k_B T, \nonumber \\
    \text{SNR}_{\text{cell}}(\omega) = \frac{\tau_{\text{leak}}}{8 \pi R^2 c k_B T} S^I(\omega).
\end{gather}
Using this measurement, the cell can achieve a voltage resolution of $\sim 0.25$ mV within 10 $\mu$s for slow signals, with resolution improving as signals get faster.

\section{Collective sensing}
We now consider that the signal is sensed by $N$ ``receiver'' ion channels located at coordinates ${\bf y}_1, \ldots, {\bf y}_N$ on the membrane. We assume the ion channels are separated by a distance much larger than $l_D$. The measurement $M_j(t)$ made by ion channel $j$ is the charge around its location ${\bf y}_j$. This communication scheme is shown in Figure \ref{fig:spherical_soma}. Signal transmission and the shared Johnson–Nyquist noise are dominated by the zeroth harmonic; higher-harmonic contributions are local and short-ranged (see Supplemental Material for details). Therefore, each channel would sense the same signal in the absence of noise, and this signal would be proportional to the average charge. Since signals are proportional to channel area, we conclude that $\langle M_j(\omega) \rangle = \frac{\sigma^2}{2R^2} \langle M_{\text{cell}}(\omega) \rangle$. Therefore, the single-channel input response kernel is $\chi^{MI}_{\text{channel}}(\omega) = \frac{\sigma^2}{2R^2} \chi^{MI}_{\text{cell}}(\omega)$.

As we showed in Section \ref{sec:noise}, each channel experiences Johnson-Nyquist noise and shot noise. Since shot noise comes from local charge fluctuations, it has the same form as in the infinite-membrane case. We obtain the Johnson-Nyquist fluctuations from our continuum charge dynamics (see Supplemental Material for details). Due to the dominance of the zeroth harmonic, Johnson-Nyquist fluctuations are correlated between all channels. Therefore, the channel measurements have the structure
\begin{align}
    M_j(\omega) = \chi^{MI}_{\text{channel}}(\omega) I(\omega) + \xi_{\text{JN}}(\omega) + \xi_{\text{shot}}^j(\omega),
\end{align}
where the shot noise sources $\xi_{\text{shot}}^j(\omega)$ are uncorrelated with each other, and are also uncorrelated with the Johnson-Nyquist noise $\xi_{\text{JN}}(\omega)$. Since the Johnson-Nyquist fluctuations come from the zeroth harmonic, their spectrum satisfies $S^M_{\text{channel, JN}}(\omega) = \frac{\sigma^4}{4R^4} S^M_{\text{cell}}(\omega)$.

We take the measurement made by the collective of channels to be the average of the recorded measurements: $\frac{1}{N} \sum_j M_j(t)$. Given our noise structure, the signal-to-noise ratio of this communication channel is 
\begin{align}
    \text{SNR}_N(\omega) = \frac{N |\chi^{MI}_{\text{channel}}(\omega)|^2}{S^M_{\text{shot}}(\omega) + N S^M_{\text{channel, JN}}(\omega)} S^I(\omega).
\end{align}
The quality of the communication channel can also be quantified using trajectory mutual information \cite{tostevin_mutual_2009, komaee_mutual_2020}. It can be shown that the mutual information between the trajectories of $I(t)$ and the vector of measurements $(M_1(t), \ldots, M_N(t))$ can be fully characterized in terms of the signal-to-noise ratio above \cite{betancourt_mutual_nodate}.

\begin{figure}
    \centering
    \includegraphics[width = \linewidth]{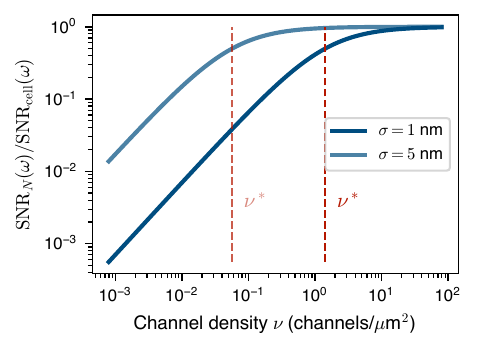}
    \caption{Saturation of the $N$-channel signal-to-noise ratio $\text{SNR}_N(\omega)$ to the whole-cell signal-to-noise ratio $\text{SNR}_{\text{cell}}(\omega)$ for the parameters in Table \ref{tab:parameters} and $\omega = \tau_{\text{leak}}^{-1}$.}
    \label{fig:snr}
\end{figure}

For a given frequency, the signal-to-noise ratio $\text{SNR}_N(\omega)$ converges to $\text{SNR}_{\text{cell}}(\omega)$ at large $N$, as shown in Figure \ref{fig:snr}. The two sources of noise naturally give rise to two regimes: one where individual shot noise dominates and one where cross-correlations from Johnson-Nyquist noise dominate. We can identify the point of saturation with a critical number of channels $N^*(\omega)$ that satisfies $S^M_{\text{shot}}(\omega) = N^*(\omega) S^M_{\text{channel,JN}}(\omega)$. Since $N^*(\omega) \propto R^2$, it is convenient instead to characterize a saturation channel density $\nu^*(\omega) \coloneqq N^*(\omega)/4\pi R^2$ From our previous characterization of noise, this critical density is
\begin{align}
    \nu^*(\omega) &= \frac{\alpha \tau_D^2}{32 \pi c l_D \tau_{\text{leak}} \sigma^2} (1+\omega^2 \tau_{\text{leak}}^2) \nonumber \\
    &\eqqcolon \nu_0^* (1+\omega^2 \tau_{\text{leak}}^2).
\end{align}
As opposed to \cite{berg_physics_1977}, the number of receptors at which sensing saturates grows quadratically with $R$, instead of linearly. Using the parameters in Table \ref{tab:parameters}, the number of channels per unit area needed to sense slow signals is $\nu_0^* \approx 10^{-2}-10^0$ channels/$\mu$m$^2$, while for $\omega^{-1} \sim 10$ $\mu$s the required density is $10^2 - 10^4$ channels/$\mu$m$^2$.

The signal-to-noise saturation result can be understood intuitively from the structure of the noise. For $N\ll N^*(\omega)$, increasing the number of channels linearly increases the signal-to-noise ratio. This is because different channels can effectively make independent signal measurements, reducing the effect of thermal noise in the integrated measurement. For $N \gg N^*(\omega)$, the channels are limited by the global Johnson-Nyquist fluctuations, and more information cannot be extracted from the signal. In this regime, the effect of cross-correlations becomes so large that the receivers essentially obtain the same amount of information as if it they were measuring the total charge distribution of the sphere.
\section{Discussion}
In this paper, we have shown that electrical communication mediated by ion channels is subject to thermal fluctuations that manifest in two forms: shot noise and Johnson-Nyquist noise. Shot noise dominates at the single-channel level, and sets an upper bound on sensory sensitivity.  Real channels have been shown to sense voltage with an accuracy that is within an order of magnitude of our bound, suggesting that this noise source is limiting.  At larger scales, Johnson-Nyquist dominates and sets a bound on the accuracy with which an applied voltage, e.g. from a dendrite, can be estimated by a perfect instrument that can make a readout of the entire cell's charge distribution.  For a neuron which must estimate the voltage from a large number of ion channels, we find that the accuracy of the perfect instrument can be approached with a surprisingly small number of channels.  Our analysis gives us absolute bounds on electrical communication based on the physics of thermal fluctuations, resembling the results of Berg and Purcell \cite{berg_physics_1977}.  However, our results do differ in interesting ways that arise from the different structure of thermal fluctuations in charge density. Berg and Purcell found that the density of receptors needed to achieve near perfect accuracy was inversely proportional to cell size. Instead we find that a fixed but small and frequency dependent fraction of the cell surface must be covered in ion channels, regardless of cell size. 

In this analysis, we made several simplifying assumptions about how ion channels sense their local environment. We abstracted from the details of real channels and assumed they made simple measurements of bulk charge.  Our results serve as a bound for an ion channel which operates by measuring local charge density. As with chemical sensing, details of implementation likely add order one prefactors that diminish accuracy of schemes that are otherwise optimal \cite{bialek_physical_2005, kaizu_berg-purcell_2014}.  However, the structure of electrical signaling is more complex, and potentially allows for a qualitatively different sensing scheme to overcome our shot noise bound.  In electrical communication, all signal is sent through surface electrical modes, but local measurements of charge density are corrupted by shot noise.  Because electrical fields are long-ranged, it is in principle possible to measure electric fields without measuring charged particle density.  In real channels, voltage-gating is a complicated process modulated by charged residues \cite{catacuzzeno_70year_2022, noda_primary_1984, sands_voltage-gated_2005, catterall_ion_2010}.  In a variety of sodium channels, a charged helix near the pore slides up and down in response to applied voltages \cite{catterall_molecular_1986, yang_evidence_1995, payandeh_crystal_2011, wisedchaisri_resting-state_2019}, while in KvAP potassium channels, a charged paddle moves away from the membrane \cite{jiang_x-ray_2003, jiang_principle_2003, ruta_functional_2003}.  It is unclear whether these schemes are best summarized as measuring charge separation.  In the supplement we consider an alternative sensor consisting of a charged plate in the membrane whose displacement reads out the membrane voltage. We find that while the contributions from Johnson noise are shared, local fluctuations take a qualitatively different form than the shot noise we found for a charged particle counting sensor. We cannot presently show that it is impossible to build a sensor which is less local noise than the shot noise contribution. 

Still, it is striking that the shot noise arising from particle counting seems to capture the sensitivity of real channels.  The most direct measurements of channel gating come from single channel patch clamp experiments.  Sodium channels have been shown to respond in less than a few hundreds of $\mu$s to voltage changes as small as 10 mV \cite{aldrich_reinterpretation_1983, kriegeskorte_cold_2023}, a factor of 10 from the bound we derived by considering the need to be statistically significant over shot noise.  In vivo, sodium  channels that mediate the action potential in nodes of Ranvier must be at least an order of magnitude faster, responding in less than 10$\mu$s.  Saltatory conduction from one node to the next takes $10\mu$s, but this includes the time for electrical signals to propagate from one node to the next.  While a full understanding of how thermal fluctuations limit this transduction modality is beyond the scope of this manuscript, these simple considerations suggest that the ion channels which mediate conduction may be near the limits of sensitivity set by shot noise.   

Our calculations also make predictions for the expected density of voltage sensitive channels in neuronal membranes, above which measurements are effectively the same as the perfect instrument. We predict that achieving this perfect performance requires just .01-1 channels per square micron for sensors which can make measurements on timescales slower than the RC time of the neuron, typically in the range of ms. Despite using a simplified geometry of neuronal electrical communication, our predictions are in the range of channel densities observed in real experiments \cite{lorincz_molecular_2010, battefeld_heteromeric_2014, hu_supercritical_2014}. However, we also predict that channels which must measure voltage on faster timescales require a density which scales quadratic in the measurement speed.  For channels in the nodes of Ranvier, where sensing must occur in $10$ $\mu$s, this predicts a four order of magnitude increase in density.  Indeed, channel density in the axon initial segment and in the nodes of Ranvier is on the order of hundreds of channels per $\mu$m$^2$ \cite{kole_action_2008, lorincz_molecular_2010, battefeld_heteromeric_2014}. It is possible that in some instances channel density is instead set by other factors.  In particular, voltage gated ion channels play dual roles in sensing and communication, and channel density must also be sufficient to mediate a sufficient current to downstream processes.

\section{Acknowledgments}
We thank Isabella Graf, Michael Abbott, Matthew Leighton, Asheesh Momi and Yu Fu for useful comments on the manuscript.

This work was supported by the NIH award R35 GM138341 (B.B.M.), Sloan Foundation grant no. G-2023-19668 (B.B.M.), and the Yale Program in Physical and Engineering Biology (J.M.B.).

\bibliography{ion_channel_sensitivity}
\bibliographystyle{apsrev4-2}

\clearpage
\onecolumngrid
\appendix

\makeatletter
\renewcommand{\appendixname}{}
\def\@seccntformat#1{\csname the#1\endcsname.\quad}
\renewcommand\section{
  \@startsection{section}{1}{\z@}
  {2.5ex \@plus 1ex \@minus .2ex}
  {1.5ex \@plus .2ex}
  {\large\bfseries}
}
\renewcommand\subsection{
  \@startsection{subsection}{2}{\z@}
  {2.0ex \@plus 0.8ex \@minus .2ex}
  {1.0ex \@plus .2ex}
  {\normalsize\bfseries}
}
\makeatother

\newtheorem{proposition}{Proposition}
\newtheorem{definition}{Definition}
\newtheorem{lemma}{Lemma}
\newtheorem{conjecture}{Conjecture}
\newtheorem{assumption}{Assumption}
\newtheorem{theorem}{Theorem}

\setcounter{equation}{0}
\setcounter{figure}{0}
\setcounter{table}{0}
\setcounter{page}{1}
\setcounter{section}{0}

\makeatletter
\renewcommand{\theequation}{\thesection.\arabic{equation}}
\renewcommand{\thefigure}{S\arabic{figure}}
\renewcommand{\thesection}{S\arabic{section}}
\renewcommand{\tocname}{Table of Contents for SM}
\makeatother

\begin{center}
\textbf{\large Supplemental Material to\\ 
Thermal fluctuations set fundamental limits on ion channel function} \\
\vspace{10pt}

\vspace{10pt}
Jose M. Betancourt and Benjamin B. Machta \\
\textit{
Department of Physics, Yale University
}\\
\textit{Quantitative Biology Institute, Yale University, New Haven, CT 06520
}
\end{center}

\vspace{10pt}

\addtocontents{toc}{\protect\setcounter{tocdepth}{2}}
\revtextableofcontents

\clearpage

\section{Introduction}
In this Supplemental Material, we show the explicit derivations for the expressions shown in the main text. Specifically, we compute the noise statistics for shot noise and Johnson-Nyquist noise on both a flat and a spherical membrane. We also characterize the communication channel in the spherical case.

\subsection{Notation}
Throughout this document, we will use the following notation:
\begin{itemize}
    \item Vectors ${\bf r}$ represent arbitrary coordinates in $\mathbb{R}^3$, while vectors ${\bf x}$ represent coordinates on the surface of the membrane (for both the flat and spherical membrane).
    \item The element $ds({\bf x})$ represents the area element at the point ${\bf x}$ on the membrane. For example, if using spherical coordinates for the spherical membrane of radius $R$, then $ds({\bf x}) = R^2 \sin(\theta) \, d\theta \, d\phi$.
    \item The operator $\partial_n$ denotes the directional derivative in the direction of the normal to the membrane. We take the convention of the normal to be $\hat{{\bf z}}$ for the planar membrane and $\hat{{\bf r}}$ for the spherical one.
    \item Fourier transforms, in both space and time, are implicit. This means that $f(\omega)$ is understood to be the Fourier transform of $f(t)$. We adopt the following convention for the time transform:
    \begin{align*}
        f(\omega) = \int dt \, f(t) e^{-i\omega t}, \quad f(t) = \frac{1}{2 \pi} \int d\omega \, f(\omega) e^{i \omega t},
    \end{align*}
    and similarly for the spatial transform. When transforming spatial coordinates, we explicitly state if only a subset of coordinates are being transformed.
    \item We use spherical harmonics for the analysis of the spherical membrane. We denote these with $Y_\ell^m({\bf x})$, which corresponds to the value of the $(\ell, m)$ harmonic evaluated at the point ${\bf x}/R$ on the unit sphere. For a function $g({\bf x})$, we define its harmonic components as
    \begin{align*}
        g_\ell^m = \frac{1}{R^2} \int ds({\bf x}) \, g({\bf x}) Y_\ell^{m*}({\bf x}).
    \end{align*}
\end{itemize}

\section{Charge dynamics in the Poisson-Nernst-Planck model}
\subsection{Charge dynamics in the bulk}
The electrical dynamics in the bulk are determined by the movement of charged ions. We use the Poisson-Nernst-Planck (PNP) model to describe the dynamics of these ions. We assume there are $M$ ionic species, and that the state of the system is characterized by their concentrations $n_1({\bf r}), \ldots, n_M({\bf r})$. We begin by specifying the free energy functional
\begin{align} \label{eq:original_free_energy}
    {\cal F}_0[{\bf n}] = U[{\bf n}] + \frac{1}{\beta} \sum_i \int d^3 {\bf r} \, n_i({\bf r}) \log(n_i({\bf r})/n_0),
\end{align}
where the first term is the potential energy from electrostatic interactions and the second term is an entropic contribution. The dynamics of this model are characterized by the transport of ions: 
\begin{align}
    \partial_t n_i + \nabla \cdot {\bf j}_i = 0, \quad {\bf j}_i = -\beta D n_i \nabla \left( \frac{\delta {\cal F}_0}{\delta n_i} \right).
\end{align}
Here $D$ is the diffusion constant, which we assume to be the same for all ions for simplicity. These result in
\begin{align} \label{eq:ionic_current}
    {\bf j}_i = - D \nabla n_i + \beta D n_i z_i e {\bf E}.
\end{align}

The net charge in the bulk is generated from deviations from the average bulk concentrations ${\bf n}^0$. These satisfy $\sum_i n_i^0 e z_i = 0$ because of neutrality, where $z_i$ is the valence of species $i$. Specifically, we can write $n_i = n_i^0 + \hat{n}_i$. We can define the local density fluctuation and total current as
\begin{align}
    \rho({\bf r},t) \coloneqq \sum_i e z_i \hat{n}_i, \quad {\bf J} \coloneqq \sum_i e z_i {\bf j}_i.
\end{align}
Note that he net charge fluctuations satisfy the continuity equation
\begin{align}
    \partial_t \rho + \nabla \cdot {\bf J} = 0.
\end{align}
Since we consider small fluctuations, we have $\hat{n}_i \ll n_i^0$, and the total current is
\begin{align} \label{eq:charge_current}
    {\bf J} = \alpha {\bf E} - D \nabla \rho,
\end{align}
where the effective conductance is given by
\begin{align}
    \alpha \coloneqq \beta D \sum_i n_i^0 e^2 z_i^2.
\end{align}

To close the dynamics of the charge density, we use the fact that the electric field satisfies Gauss's law
\begin{align}
    \nabla \cdot {\bf E} = \frac{\rho}{\epsilon},
\end{align}
where $\epsilon$ is the permittivity of the bulk. The continuity equation, then, becomes
\begin{align} \label{eq:density_dynamics}
    \tau_D \partial_t \rho = - \rho + l_D^2 \nabla^2 \rho,
\end{align}
where the Debye time and the Debye length are, respectively, defined by
\begin{align} 
    \tau_D \coloneqq \frac{\epsilon}{\alpha}, \quad l_D^2 \coloneqq \frac{D \epsilon}{\alpha}.
\end{align}
In the cellular environment, we have $\tau_D \approx 10^{-9}$ s and $l_D \approx 7 \times 10^{-10}$ m.

To build intuition, we can analyze this equation in the absence of a membrane. Transforming this equation in space, we obtain
\begin{align}
    \partial_t \rho({\bf k}) = - \frac{(1 + k^2 l_D^2)}{\tau_D} \rho({\bf k}).
\end{align}
This means that long wavelength modes (with $k l_D \ll 1$) decay uniformly with timescale $\tau_D$, so charge gets quickly screened at distances greater than $l_D$.

\subsection{Electric potential in the presence of a membrane} \label{subsec:membrane_potential}
We now analyze the effect of adding a membrane on the dynamics of the charge density. We model the membrane as a slab in the region $|z| < d/2$ with permittivity $\epsilon_m < \epsilon$. 

First, we determine the potential generated by an arbitrary charge density profile. This will allow us to characterize the dynamics near the membrane. The potential solves the Poisson equation both inside and outside the membrane. Since there is no charge inside the membrane, this can be written as
\begin{align}
    \nabla^2 \Phi({\bf r}) =
    \begin{cases}
        -\frac{\rho({\bf r})}{\epsilon} & \textrm{if } |z| > d/2, \\
        0 & \textrm{if } |z| < d/2.
    \end{cases}
\end{align}
To simplify the distinction between the regions, we denote the potential with $\Phi^+$ in the region $z > d/2$, with $\Phi^-$ in the region $z< -d/2$ and with $\Phi^m$ in the region $|z| < d/2$. At the interfaces between the membrane and the bulk, the normal component of the electric displacement field must be continuous. This yields the interface conditions (abusing notation, here ${\bf x}$ is a 2D vector representing coordinates on the membrane)
\begin{align} \label{eq:interface_conds}
    \epsilon \partial_n\Phi^+ ({\bf x}, d/2) &= \epsilon_m \partial_n \Phi^m({\bf x},d/2) \coloneqq \lambda^+({\bf x}), \nonumber \\
    \epsilon \partial_n\Phi^- ({\bf x}, -d/2) &= \epsilon_m \partial_n \Phi^m({\bf x},-d/2) \coloneqq \lambda^-({\bf x}).
\end{align}
The fields $\lambda^\pm$ simplify the notation when characterizing the solution. Additionally, the potential must be continuous at the boundaries:
\begin{align} \label{eq:potential_continuity}
    \Phi^+({\bf x},d/2) = \Phi^m({\bf x},d/2), \quad \Phi^-({\bf x},-d/2) = \Phi^m({\bf x},-d/2).
\end{align}
To solve for the potential, we transform the coordinates parallel to the membrane. With this, the Poisson equation takes the form
\begin{align}
    (\partial_z^2 - k^2) \Phi({\bf k},z) =
    \begin{cases}
        -\frac{\rho({\bf k},z)}{\epsilon} & \textrm{if } |z| > d/2, \\
        0 & \textrm{if } |z| < d/2.
    \end{cases}
\end{align}
Along with the boundary conditions, this fully determines the potential. For the bulk, we have
\begin{align} \label{eq:bulk_potential}
    \Phi^+({\bf k},z) &= - \frac{\lambda^+({\bf k})}{\epsilon k} e^{-k(z-d/2)} + \frac{1}{2\epsilon k} \int_{d/2}^\infty dz' \, \rho({\bf k},z') \left[ e^{-k|z-z'|} + e^{-k(z+z'-d)} \right], \nonumber \\
    \Phi^-({\bf k},z) &= \frac{\lambda^-({\bf k})}{\epsilon k} e^{k(z+d/2)} + \frac{1}{2\epsilon k} \int_{-\infty}^{-d/2} dz' \, \rho({\bf k},z') \left[ e^{-k|z-z'|} + e^{k(z+z'+d)} \right].
\end{align}
Inside the membrane, the potential is
\begin{align} \label{eq:membrane_potential}
    \Phi^m({\bf k},z) = \frac{1}{2 \epsilon_m k} \left[ \frac{\cosh(kz)}{\sinh(kd/2)} (\lambda^+({\bf k}) - \lambda^-({\bf k})) + \frac{\sinh(kz)}{\cosh(kd/2)} (\lambda^+({\bf k}) + \lambda^-({\bf k})) \right].
\end{align}
The fields $\lambda^\pm$ are then determined by invoking continuity. For this, it is useful to define the quantities
\begin{align}
    \langle \rho^+({\bf k}) \rangle \coloneqq k \int_{d/2}^\infty dz' \, \rho({\bf k},z') e^{-k(z-d/2)}, \quad \langle \rho^-({\bf k}) \rangle \coloneqq k \int_{-\infty}^{-d/2} dz' \, \rho({\bf k},z') e^{k(z+d/2)}.
\end{align}
These are exponentially-weighted averages of the charge density for each mode. Instead of characterizing the fields individually, it is more useful to work with the following quantities:
\begin{align} \label{eq:lambdas}
    \lambda^+({\bf k}) + \lambda^-({\bf k}) &= \frac{1}{k} \left\{1 + \frac{\epsilon}{\epsilon_m} \tanh(kd/2) \right\}^{-1} (\langle \rho^+({\bf k}) \rangle - \langle \rho^-({\bf k}) \rangle), \nonumber \\
    \lambda^+({\bf k}) - \lambda^-({\bf k}) &= \frac{\epsilon_m}{\epsilon k} \frac{\tanh(kd/2)}{1 + \frac{\epsilon_m}{\epsilon} \tanh(kd/2)} (\langle \rho^+({\bf k}) \rangle + \langle \rho^-({\bf k}) \rangle).
\end{align}
This will allow us to obtain the appropriate boundary conditions for the charge dynamics.

\subsection{Charge profile boundary conditions}
In the bulk the charge dynamics evolve according to Eq. \eqref{eq:density_dynamics}. We assume there is no free charge in the membrane and that ions cannot cross the boundary into the membrane. This enforces a no-flux condition on all ionic species:
\begin{align} \label{eq:no_flux}
    j^\perp_i({\bf x},z=0^+) = j^\perp_i({\bf x},z=0^-) = 0,
\end{align}
where $j^\perp_i$ is the component of the current perpendicular to the membrane. Here we have abused notation further by only considering $z$ coordinates in the bulk, such that $z=0^+$ correspond to the top of the membrane and $z=0^-$ to the bottom. This simplifies the calculations, since the effects of the interior of the membrane only matter through the fields $\lambda^\pm$ in the rest of the analysis.

Using Eq. \eqref{eq:ionic_current} for the currents, the no-flux conditions translate to Neumann conditions for the ionic concentrations:
\begin{align}
    -D \partial_z n_i({\bf x},z=0^+) + \beta D n_i z_i e E^\perp({\bf x},z=0^+) = -D \partial_z n_i({\bf x},z=0^-) + \beta D n_i z_i e E^\perp({\bf x},z=0^-) = 0.
\end{align}
Summing over species and transforming the ${\bf x}$ coordinates, we obtain analogous boundary conditions for the charge density:
\begin{align}
    \alpha E^\perp({\bf k},z=0^+) - D \partial_z \rho({\bf k},z=0^+) = \alpha E^\perp({\bf k},z=0^-) - D \partial_z \rho({\bf k},z=0^-) = 0.
\end{align}
Using our solution for the potential we can get the electric fields at the membrane:
\begin{align}
    E^\perp({\bf k}, z=0^+) &= -\partial_z \Phi^+({\bf k}, z=0^+) = - \frac{\lambda^+({\bf k})}{\epsilon}, \nonumber \\
    E^\perp({\bf k}, z=0^-) &= -\partial_z \Phi^-({\bf k}, z=0^-) = - \frac{\lambda^-({\bf k})}{\epsilon}.
\end{align}

Now, the boundary conditions for the charge density can also be expressed using the following linear combinations:
\begin{align} \label{eq:boundary_conditions}
    \partial_z \rho({\bf k}, z=0^+) + \partial_z \rho({\bf k}, z=0^-) &= -\frac{\alpha}{\epsilon D} [\lambda^+({\bf k}) + \lambda^-({\bf k})], \nonumber \\
    \partial_z \rho({\bf k}, z=0^+) - \partial_z \rho({\bf k}, z=0^-) &= -\frac{\alpha}{\epsilon D} [\lambda^+({\bf k}) - \lambda^-({\bf k})].
\end{align}
This is useful since an arbitrary profile $\rho({\bf k},z)$ can be decomposed into an odd and an even part, each of which satisfies one of these boundary conditions automatically. Explicitly:
\begin{align}
    \rho_{\text{odd}}({\bf k},z) = \frac{\rho({\bf k},z) - \rho({\bf k},-z)}{2}, \quad \rho_{\text{even}}({\bf k},z) = \frac{\rho({\bf k},z) + \rho({\bf k},-z)}{2}.
\end{align}
In this decomposition, $\rho_{\text{odd}}$ always satisfies the second boundary condition in Eq. \eqref{eq:boundary_conditions}, while $\rho_{\text{even}}$ always satisfies the first one. When characterizing measurements, we will focus on charge differences between the two sides of the membrane, which are only affected by $\rho_{\text{odd}}$. Therefore, from now on we only focus on odd profiles.

If our profile is odd, we can specify it for the half-interval $z>0$ to obtain the full solution. Furthermore, we only need to have it satisfy the first boundary condition in Eq. \eqref{eq:boundary_conditions} to satisfy no-flux. Using Eq. \eqref{eq:lambdas}, our boundary condition can be written as (recall that $l_D^2 = D\epsilon/\alpha$)
\begin{align}
    \partial_z \rho({\bf k},z=0^+) = - \frac{1}{l_D^2} \left\{1 + \frac{\epsilon}{\epsilon_m} \tanh(kd/2) \right\}^{-1} \int_0^\infty dz' \, \rho({\bf k}, z') e^{-kz'}.
\end{align}
This boundary condition, along with the condition $\rho({\bf k},z\to \infty) \to 0$, will be sufficient to determine the structure of the dynamics of the charge profile.

\subsection{Mode structure of the charge dynamics} \label{subsec:membrane_dynamics}
To simplify exposition, we will fix ${\bf k}$ throughout this subsection and omit the dependence of $\rho$ on ${\bf k}$. Additionally, we define $\psi \coloneqq \left\{1 + \frac{\epsilon}{\epsilon_m} \tanh(kd/2) \right\}^{-1}$. Note that $\psi < 1$ and, for long wavelengths,
\begin{align}
    \psi \approx 1 - \frac{\epsilon}{2 \epsilon_m} kd.
\end{align}

We want to characterize the dynamics of the charge density profile, which follows the PDE
\begin{align}
    \tau_D \partial_t \rho = - (1 + k^2 l_D^2) \rho + l_D^2 \partial_z^2 \rho,
\end{align}
subject to the boundary condition
\begin{align}
    \partial_z \rho(z=0^+) = - \frac{\psi}{l_D^2} \int_0^\infty dz' \, \rho(z') e^{-kz'}.
\end{align}
To simplify the analysis, we can write the problem in terms of the following dimensionless variables:
\begin{align}
    s \coloneqq (1+k^2 l_D^2) \frac{t}{\tau_D}, \quad \zeta \coloneqq (1+k^2 l_D^2)^{1/2} \frac{z}{l_D}, \quad \kappa \coloneqq (1+k^2 l_D^2)^{-1/2} kl_D.
\end{align}
Our PDE, then, becomes
\begin{align}
    \partial_s \rho = -\rho + \partial_\zeta^2 \rho,
\end{align}
with corresponding boundary condition
\begin{align}
    \partial_\zeta \rho(\zeta = 0^+) = - \frac{\psi}{(1+kl_D^2)} \int_0^\infty d\zeta' \, \rho(\zeta') e^{-\kappa \zeta'}.
\end{align}

We want to find the modes of the charge dynamics above, which are the eigenfunctions of the operator $-1 + \partial_\zeta^2$ that satisfy the boundary condition. It is convenient to characterize these modes by their dimensionless relaxation time, which is the reciprocal of their corresponding eigenvalue. We denote with $\phi_{s_0}$ the mode with dimensionless relaxation time $s_0$. By definition, $\phi_{s_0}$ satisfies
\begin{align}
    - \frac{1}{s_0} \phi_{s_0}(\zeta) = - \phi_{s_0}(\zeta) + \phi_{s_0}''(\zeta).
\end{align}
The general solution to this ODE is
\begin{align}
    \phi_{s_0}(\zeta) = A \exp\left[ \left( \frac{s_0 - 1}{s_0} \right)^{1/2} \zeta \right] + B \exp\left[ -\left( \frac{s_0 - 1}{s_0} \right)^{1/2} \zeta \right].
\end{align}
We can clearly see that there are two very different regions: when $s_0 > 1$ we have exponentially decaying modes (in space) and when $s_0 < 1$ we have oscillating modes. Let us go through these cases separately.

\subsubsection{Slow timescales ($s_0 > 1$)}
Since we do not want exponentially increasing solutions, we must have $A=0$, and can set $B=1$. For this mode, the boundary condition translates to
\begin{align}
    \left( \frac{s_0 - 1}{s_0} \right)^{1/2} = \frac{\psi}{(1+kl_D^2)} \left\{ \kappa + \left( \frac{s_0 - 1}{s_0} \right)^{1/2} \right\}^{-1}.
\end{align}
This equation has a unique solution $s_0^*$. We can obtain the (dimensionful) decay timescale of this mode:
\begin{align}
    \tau_{\text{slow}} \coloneqq \frac{\tau_D}{(1+k^2 l_D^2)} s_0^*.
\end{align}
For $k \ll l_D$, this decay time is
\begin{align}
    \tau_{\text{slow}} \approx \frac{2}{\alpha k} \left( \frac{1}{c_0} + \frac{2l_D}{\epsilon} \right)^{-1},
\end{align}
where $c_0 = \epsilon_m/d$ is the capacitance per unit area of the membrane. These is precisely the timescale of the coarse-grained 2D dynamics in \cite{bryant_physical_2023}, where the capacitance is replaced by an effective capacitance $\frac{1}{c_{\text{eff}}} = \frac{1}{c_0} + \frac{2l_D}{\epsilon}$.

The decay lengthscale for the slow mode is
\begin{align}
    \left( \frac{s_0^*}{s_0^* - 1} \right)^{1/2} \frac{l_D}{(1_k^2 l_D^2)^{1/2}} \approx l_D.
\end{align}
Since communication usually takes place at distances much larger than the Debye length, we see that the only long-lived mode that can carry information is this slow mode. All the other modes decay at a timescale smaller than $\tau_D$ in this regime.

\subsubsection{Fast timescales ($s_0 < 1$)}
For this case, we can write our modes as
\begin{align}
    \phi_{s_0}(\zeta) = \sin\left[ \left( \frac{1 - s_0}{s_0} \right)^{1/2} \zeta \right] + Q(s_0) \cos\left[ \left( \frac{1 - s_0}{s_0} \right)^{1/2} \zeta \right].
\end{align}
The boundary condition takes the form
\begin{align}
    \left( \frac{1 - s_0}{s_0} \right)^{1/2} = -\frac{\psi}{(1+k^2l_D^2)} \left[ \frac{ Q(s_0) \kappa + \left( \frac{1 - s_0}{s_0} \right)^{1/2}}{ \kappa^2 + \left( \frac{1 - s_0}{s_0} \right)} \right].
\end{align}
This condition can always be satisfied by choosing
\begin{align}
    Q(s_0) = - \frac{1}{\kappa} \left(\frac{1-s_0}{s_0}\right)^{1/2} \left[ \frac{\kappa^2 + \left(\frac{1-s_0}{s_0}\right)}{\psi/(1+k^2 l_D^2)} + 1 \right].
\end{align}
This means that there is a continuum of oscillatory ``fast modes'' which decay at a timescale shorter than the Debye time.
\section{Measurement fluctuations} \label{sec:measurement_fluctuations}
\subsection{Coupling of the measurement to the free energy}
We consider an ion channel that performs the following measurement of the charge in the bulk:
\begin{align}
    M(t) = \int d^2{\bf x} \, dz \, e^{-\lVert {\bf x} \rVert^2/2\sigma^2} e^{-|z|/h} \text{sgn}(z) \rho({\bf x},z,t).
\end{align}
Since this measurement is a functional of the charge density, we can consider a perturbation of the free energy that takes the same form as a coupling of the measurement to a fictitious field $\zeta$:
\begin{align}
    {\cal F}[{\bf n},\zeta] = {\cal F}_0[{\bf n}] - \zeta M.
\end{align}

This coupling changes the dynamics of the model by altering the currents obtained from Eq. \eqref{eq:ionic_current}. Let $\chi^{M \zeta}(\omega)$ denote the linear response kernel of the measurement to the coupling $\zeta$, defined by
\begin{align}
    M(\omega) = \chi^{M\zeta}(\omega) \zeta(\omega).
\end{align}
According to the fluctuation-dissipation theorem, the spectrum of equilibrium fluctuations of $M$ satisfies
\begin{align}
    S^M(\omega) = - \frac{2 k_B T}{\omega} \text{Im}[\chi^{M \zeta}(\omega)].
\end{align}
Therefore, we can obtain the spectrum of the fluctuations by computing the response kernel $\chi^{M\zeta}(\omega)$. This is done in this section.

\subsection{Perturbed dynamics and boundary conditions}
Let us define the kernel
\begin{align}
    \Gamma({\bf r}) \coloneqq e^{-\lVert {\bf x} \rVert^2/2\sigma^2} e^{-|z|/h} \text{sgn}(z).
\end{align}
The ionic currents from the perturbed free energy can be written as
\begin{align}
    {\bf j}_i = -D \nabla n_i + \beta D n_i z_i e {\bf E} + \zeta \beta n_i z_i e \nabla \Gamma.
\end{align}
Summing over species, we get the modified charge current
\begin{align}
    {\bf J} = \alpha {\bf E} - D \nabla \rho + \alpha \zeta \nabla \Gamma.
\end{align}
Using conservation of chage, this leads to the modified dynamics
\begin{align}
    \tau_D \partial_t \rho = - \rho + l_D^2 \nabla^2 \rho + \epsilon \zeta \nabla^2 \Gamma.
\end{align}
The Laplacian of $\Gamma$ is
\begin{align}
    \nabla^2 \Gamma = \textrm{sgn}(z) \left[ \frac{1}{h^2} + \frac{\lVert {\bf x} \rVert^2}{\sigma^4} - \frac{2}{\sigma^2} \right] e^{-\lVert {\bf x} \rVert^2/2\sigma^2} e^{-|z|/h}.
\end{align}
Therefore, taking the Fourier transform of the ${\bf x}$ coordinates, we obtain the following dynamics in Fourier space:
\begin{align}
    \tau_D \partial_t \rho = - (1 + k^2 l_D^2) \rho + l_D^2 \partial_z^2 \rho - 2 \pi \epsilon \zeta \frac{\sigma^2}{h^2} (1 - k^2 h^2) \text{sgn}(z) e^{-k^2 \sigma^2/2} e^{-|z|/h}.
\end{align}

Going back to the modified currents, we see that the no-flux condition now reads
\begin{align}
    \partial_z \rho({\bf x}, 0^\pm) = \frac{\alpha}{D} E^\perp({\bf x}, 0^{\pm}) + \frac{\alpha \zeta}{D} \partial_z \Gamma({\bf x}, 0^{\pm}).
\end{align}
Evaluating the derivative of the kernel yields
\begin{align}
    \partial_z \Gamma({\bf x}, 0^{\pm}) = - \frac{1}{h} e^{-\lVert {\bf x} \rVert^2/2\sigma^2}.
\end{align}
This form of the perturbation only affects the odd part of the charge profile. Since $M$ only depends on the odd part of the profile, we only need to find how this part responds to the coupling $\zeta$. Therefore, we can focus on odd profiles that satisfy the following boundary condition:
\begin{align}
    \partial_z \rho({\bf x}, 0^+) + \partial_z \rho({\bf x}, 0^+) = \frac{\alpha}{D} (E^\perp({\bf x}, 0^+) + E^\perp({\bf x}, 0^-)) - 2\frac{\alpha \zeta}{D h} e^{-\lVert {\bf x} \rVert^2/2\sigma^2}.
\end{align}
Again using the solution for Poisson's equation, we can write the boundary condition for the solution on the half-plane $z>0$ as
\begin{align}
    \partial_z \rho({\bf k},0^+) = -\frac{\psi({\bf k})}{l_D^2} \int_0^\infty dz \, e^{-kz} \rho({\bf k},z) - 2 \pi \sigma^2 \frac{\alpha \zeta}{D h} e^{-k^2 \sigma^2/2}.
\end{align}

\subsection{The charge response kernel}
To find the kernel $\chi^{M \zeta}(\omega)$, we need to first determine how $\rho({\bf k},\omega)$ responds to $\zeta(\omega)$. We begin by analyzing the bulk dynamics in frequency space. It is useful to define the following quantity:
\begin{align}
    q^2 \coloneqq 1 + k^2 l_D^2 + i \omega \tau_D.
\end{align}
Then the dynamics can be written as
\begin{align}
    q^2 \rho - l_D^2 \partial_z^2 \rho = - 2\pi \epsilon \zeta \frac{\sigma^2}{h^2} (1-k^2 h^2) e^{-k^2 \sigma^2/2} e^{-z/h}.
\end{align}
We can to find the profile the solves this differential equation. Since we want solutions that do not grow exponentially as $z \to \infty$, the homogeneous solution is
\begin{align}
    \rho_h(z) = A e^{-qz/l_D}.
\end{align}
The particular solution is
\begin{align}
    \rho_p(z) = - 2 \pi \zeta \frac{\sigma^2 \epsilon}{h^2} \frac{(1-k^2 h^2)}{(q^2 - l_D^2 h^2)} e^{-k^2 \sigma^2/2} e^{-z/h}.
\end{align}
Our charge density is, then, $\rho = \rho_h + \rho_p$. To determine the coefficient $A$, we can use our boundary condition. It is useful to compute the following quantities:
\begin{align}
    \partial_z \rho(0^+) &= -\frac{Aq}{l_D} + 2 \pi \zeta \frac{\sigma^2 \epsilon}{h^3} \frac{(1-k^2 h^2)}{(q^2 - l_D^2/h^2)}  e^{-k^2 \sigma^2/2}, \nonumber \\
    \int_0^\infty dz \, e^{-kz} \rho(z) &= \frac{A}{k + q/l_D} - 2 \pi \zeta \frac{\sigma^2 \epsilon}{h} \frac{(1-k h)}{(q^2 - l_D^2/h^2)} e^{-k^2 \sigma^2/2}.
\end{align}
Plugging these into our boundary condition yields
\begin{align}
    &-\frac{Aq}{l_D} + 2 \pi \zeta \frac{\sigma^2 \epsilon}{h^3} \frac{(1-k^2 h^2)}{(q^2 - l_D^2/h^2)}  e^{-k^2 \sigma^2/2}  = \nonumber \\
    &- \frac{\psi}{l_D^2} \left[ \frac{A}{k + q/l_D} - 2 \pi \zeta \frac{\sigma^2 \epsilon}{h}\frac{(1-k h)}{(q^2 - l_D^2/h^2)} e^{-k^2 \sigma^2/2}\right] - 2 \pi \zeta \frac{\sigma^2 \epsilon}{h l_D^2} e^{-k^2 \sigma^2/2}.
\end{align}
Let us define the following dimensionless quantities:
\begin{align}
    \kappa \coloneqq k l_D, \quad \eta \coloneqq \frac{h}{l_D}.
\end{align}
Solving for $A$ yields
\begin{align}
    A = - 2 \pi \zeta \frac{\sigma^2 \epsilon}{l_D \eta} \underbrace{\left\{ \frac{\psi}{\kappa + q} - q\right\}^{-1}\left[ \frac{(1-\kappa^2 \eta^2)}{(\eta^2 q^2 - 1)} - \frac{\psi (1-\kappa \eta)}{(q^2 - 1/\eta^2)} + 1 \right]}_{\coloneqq G_0}.
\end{align}
Therefore, our solution to the charge density profile is
\begin{align}
    \rho(z,\omega) = - 2 \pi \frac{\sigma^2 \epsilon}{l_D^2 \eta^2} \left[ \eta G_0 e^{-qz/l_D} + \frac{(1-\kappa^2 \eta^2)}{(q^2 - 1/\eta^2)}  e^{-z/h} \right] e^{-k^2 \sigma^2/2} \zeta(\omega)
\end{align}
This relationship allows to build the response function of the measurement.

\subsection{The response function}
We can write the measurement in frequency space as
\begin{align}
    M(\omega) = \frac{\sigma^2}{2 \pi } \int d^2{\bf k} \, dz \, e^{-k^2 \sigma^2/2} \textrm{sgn}(z) e^{-|z|/h} \rho({\bf k},z,\omega).
\end{align}
We can use the response of $\rho$ to $\omega$ to write this integral in terms of $\zeta(\omega)$:
\begin{align}
    M(\omega) = - 2 \frac{\sigma^4 \epsilon}{l_D^2 \eta^2} \left[\int d^2{\bf k} \, e^{-k^2 \sigma^2} \int_0^\infty dz \, e^{-z/h} \left( \eta G_0 e^{-qz/l_D} + \frac{(1-\kappa^2 \eta^2)}{(q^2 - 1/\eta^2)}  e^{-z/h} \right) \right] \zeta(\omega)
\end{align}
Evaluating the integral yields
\begin{align}
    M(\omega) = - 2 \frac{\sigma^4 \epsilon}{l_D^2 \eta^2} \left[\int d^2{\bf k} \,e^{-k^2 \sigma^2}  \left( \frac{\eta G_0}{q/l_D + 1/h} + \frac{l_D \eta}{2} \frac{(1-\kappa^2 \eta^2)}{(q^2 - 1/\eta^2)}  \right) \right] \zeta(\omega).
\end{align}
Therefore, our response function is
\begin{align}
    \chi^{M \zeta}(\omega) = -\frac{\sigma^4 \epsilon}{l_D} \int d^2{\bf k} \, e^{-k^2 \sigma^2} \left( \frac{2 G_0}{q \eta + 1} + \eta\frac{(1-\kappa^2 \eta^2)}{(q^2 \eta^2 - 1)}  \right).
\end{align}

\subsection{The fluctuation spectrum}
Using the fluctuation-dissipation theorem, we have that the spectrum of fluctuations of $M$ is given by
\begin{align}
    S^M(\omega) = - \frac{2 k_B T}{\omega} \textrm{Im}[\chi^{M \zeta}(\omega)].
\end{align}
Now, we are working in a regime in which $\omega \tau_D$ is much smaller than all the other quantities in the problem. Let us define the following function:
\begin{align}
    G \coloneqq \frac{2 G_0}{q \eta + 1} + \eta \frac{(1-\kappa^2 \eta^2)}{(q^2 \eta^2 - 1)}
\end{align}
Note that $\omega$ only enters $G$ through $q$. Therefore, to first order in $\omega$, we can write
\begin{align}
    G(\omega) \approx G(\omega = 0) + \left. \frac{\partial G}{\partial q}\right|_{\omega = 0} \frac{i \omega \tau_D}{2q} 
\end{align}
where $q$ is now evaluated at $\omega=0$. This means that
\begin{align}
    \textrm{Im}[G(\omega)] \approx \frac{\partial G}{\partial q} \frac{\omega \tau_D}{2q}.
\end{align}
Using this in our expression for the fluctuations yields
\begin{align}
    S^M(\omega) \approx \frac{\sigma^4 \epsilon \tau_D}{l_D} k_B T \int d^2{\bf k} \, \frac{e^{-k^2 \sigma^2}}{q} \frac{\partial G}{\partial q}
\end{align}
Since the integral is radially symmetric, we can write it as
\begin{align}
    S^M(\omega) \approx 2 \pi\frac{\sigma^4 \epsilon \tau_D}{l_D} k_B T \int_0^\infty dk \, k \frac{e^{-k^2 \sigma^2}}{q} \frac{\partial G}{\partial q}
\end{align}
In terms of the dimensionless variable $\kappa$, this is
\begin{align} \label{eq:fluctuations}
    S^M(\omega) \approx 2 \pi\frac{\sigma^4 \epsilon \tau_D}{l_D^3} k_B T \int_0^\infty d\kappa \, e^{-\kappa^2 \sigma^2/l_D^2} \left[\frac{\kappa}{q} \frac{\partial G}{\partial q} \right]
\end{align}
The integral in this expression is a complicated function of the sensor height $h/l_D$, sensor width $\sigma/l_D$ and the other parameters of the problem. In the following section we study the problem of optimal design of sensors and characterize the scaling of noise for these.
\section{Optimal sensors and noise scaling}
\subsection{Communication and sensing}
Our analysis focuses on sensors in the context of a larger communication scheme, in which they are trying to sense signals from the environment. Often these signals originate from far away, and they need to be communicated using the electrical dynamics of the system. Following \cite{bryant_physical_2023}, we study a communication scheme in which a signal is generated at a distance $r \gg \sigma$ and is sensed by our sensor. We characterize the optimal sensing height for this problem.

\begin{figure}[!ht]
    \centering
    \includegraphics[width=0.7\linewidth]{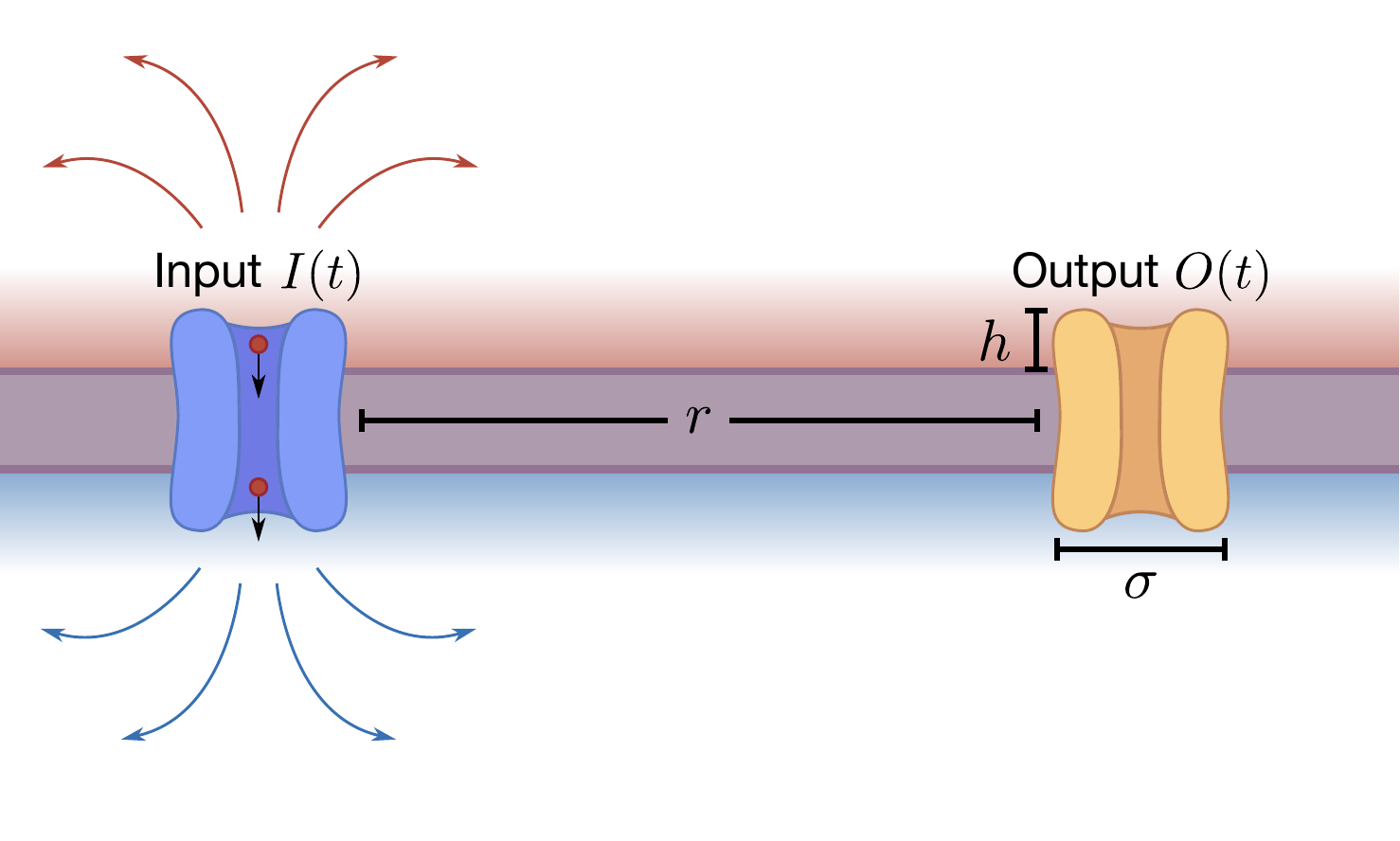}
    \caption{Communication scheme for two far away ion channels. The sender channel injects current at the membrane, while the output channel makes a local measurement of charge.}
    \label{fig:communication_scheme}
\end{figure}

Consider the communication scheme shown in Figure \ref{fig:communication_scheme}. A signal source located at a distance $r \gg \sigma$ from the sensor injects a time-dependent current $I(t)$. The output of the communication scheme is our measurement $M(t)$. As shown in \cite{bryant_physical_2023}, the information carried through this communication channel can be characterized via the following signal-to-noise ratio:
\begin{align} \label{eq:snr}
    \text{SNR}(\omega) = \frac{|\chi^{MI}(\omega)|^2 S^I(\omega)}{S^M(\omega)},
\end{align}
where $\chi^{MI}(\omega)$ is the linear response kernel of the measurement to the input and $S^I(\omega)$ is the power spectrum of the input.

The response kernel $\chi^{MI}$ depends on the dynamics of the charge density profile. For long-ranged communication, only the long-lived $z$ mode in Section \ref{subsec:membrane_dynamics} can carry information. Furthermore, as shown in \cite{bryant_physical_2023}, the main contribution to the response kernel comes from ${\bf k}$ modes where $k r$ is of order $1$, which implies $k l_D \sim l_D/r \ll 1$. As we showed in Section \ref{subsec:membrane_dynamics}, the long-lived $z$ mode decays as $e^{-|z|/l_D}$ for $k l_D \ll 1$. This means the information that can be conveyed through electrical communication is confined to a layer of width $l_D$, which is called the Debye layer. Since the relevant information is contained within the Debye layer, we can isolate the $h$ dependence of $\chi^{MI}$ as follows:
\begin{align}
    \chi^{MI}(\omega) \propto \int d^3{\bf r} \, \Gamma({\bf r}) \text{sgn}(z) e^{-|z|/l_D} \propto \left( \frac{h}{l_D + h} \right) = \left( \frac{\eta}{1 + \eta} \right).
\end{align}
Since $S^I({\omega})$ does not depend on height and we have the scaling of $\chi^{MI}(\omega)$ with $h$, we can now study how the signal-to-noise ratio behaves as a function of height, and how to use this to optimize our sensor.

\subsection{Optimal sensor height} \label{subsec:optimal_height}
\begin{figure}
    \centering
    \begin{subfigure}[t]{0.5\textwidth}
        \centering
        \includegraphics[height=0.8\textwidth]{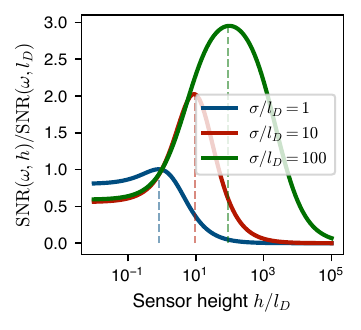}
        \caption{Dependence of the signal-to-noise ratio on sensor height for the parameters in the main text.}
    \end{subfigure}
    ~ 
    \begin{subfigure}[t]{0.5\textwidth}
        \centering
        \includegraphics[height=0.8\linewidth]{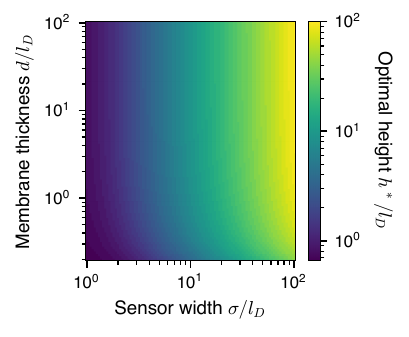}
        \caption{Optimal sensor height as a function of $\sigma$ and $d$.}
    \end{subfigure}
    \caption{Optimization of the signal-to-noise ratio for $\epsilon/\epsilon_m = 20$.}
    \label{fig:optimal_height}
\end{figure}

To obtain the optimal sensor height, we evaluate the signal-to-noise ratio in Eq. \eqref{eq:snr} as a function of height for various values of sensor width $\sigma$ and membrane height $d$. This is shown in Figure \ref{fig:optimal_height}. We see that $\text{SNR}(\omega,h)$ has a unique maximum for a wide range of parameters (Fig. \ref{fig:optimal_height}.a), so there is a well-defined optimal sensor height. The optimal sensor height seems to mostly depend on $\sigma$ rather than $d$ (Fig. \ref{fig:optimal_height}.b). 

While it may seem that there is an advantage to having a very high sensor if $\sigma$ is very large ($\sim 100$ Debye lenghts), we see that this is because $\text{SNR}(\omega,h)$ becomes very flat for high values of $\sigma$ (Fig. \ref{fig:optimal_height}.a). In fact, since noise increases with height and the signal $\chi^{MI}$ saturates at $h \sim l_D$, the sensor cannot benefit significantly from being more than a few Debye lengths high. Additionally, the fact that  $\chi^{MI}$ saturates means that in order to have a flat signal-to-noise ratio, $S^M$ must also be flat in this region. Therefore, even if the true optimal height can be large, a similar level of noise can be achieved with a sensor that is a few Debye lengths high.

\subsection{Scaling of fluctuations}
We now have a numerical characterization of the optimal height of a sensor as a function of parameters. To gain further intuition, we would like to understand what drives the scaling of the noise. As we argued in \ref{subsec:optimal_height}, signals sent from far away are confined to the Debye layer, so $l_D$ is the relevant length scale that should determine the scale of fluctuations. Therefore, we will analytically study the behavior of fluctuations for a sensor of height $h=l_D$.

Since we are interested in studying sensors that operate at length scales larger than $l_D$, we analyze what happens when we expand the fluctuations (Eq. \eqref{eq:fluctuations}) in $l_D/\sigma$. We can do this by expanding the following integrand in $\kappa$:
\begin{align}
    \frac{\kappa}{q} \frac{\partial G}{\partial q} = a_0 + a_1 \kappa + \ldots
\end{align}
This gives an expansion of the following integral:
\begin{align}
    \int_0^\infty d\kappa \, e^{-\kappa^2 \tilde{\sigma}^2} \left[\frac{\kappa}{q} \frac{\partial G}{\partial q} \right] = \tilde{a}_0 \left( \frac{l_D}{\sigma} \right) + \tilde{a}_1 \left( \frac{l_D}{\sigma} \right)^2 + \ldots,
\end{align}
where $\tilde{a}_n = a_n \int_0^\infty dx \, x^n e^{-x^2}$. Finally, we can use this to obtain an expansion of the fluctuations in $l_D/\sigma$:
\begin{align}
    S^M(\omega) = S^M_0(\omega) + S^M_1(\omega) + \ldots
\end{align}
By evaluating the terms in this expansion, we can gain an intuition for the main sources of noise in our system.

The first two coefficients in our expansion of the integrand are
\begin{align}
    a_0 = \left( 1 + \frac{1}{2} \frac{\epsilon}{\epsilon_m} \frac{d}{l_D} \right)^{-2} = \frac{4 l_D^2 c_{\text{eff}}^2}{\epsilon^2}, \quad a_1 = \frac{1}{4} \left[ 1 + \frac{3 \frac{\epsilon d}{\epsilon_m l_D}}{\left( 1 + \frac{\epsilon d}{2 \epsilon_m l_D} \right)^2} \right],
\end{align}
Plugging this into the expression for the fluctuation spectrum, we get an expansion of the form
\begin{align}
    S^M(\omega) = S^M_0(\omega) + S^M_1(\omega) + \ldots
\end{align}
With out expansion above, the leading terms in the fluctuation expansion are
\begin{align}
    S^M_0(\omega) = 4 \pi^{3/2} \frac{\sigma^3 \tau_D c_{\text{eff}}^2}{\epsilon} k_B T, \quad S^M_1(\omega) = \left[ 1 + \frac{3 \frac{\epsilon d}{\epsilon_m l_D}}{\left( 1 + \frac{\epsilon d}{2 \epsilon_m l_D} \right)^2} \right] \frac{\pi}{4} \frac{\sigma^2 \epsilon \tau_D}{l_D} k_B T.
\end{align}
From this expansion we can define two natural scales of noise:
\begin{align} \label{eq:noise_scales}
    S^M_{\text{JN}}(\omega) \coloneqq 4 \pi^{3/2} \frac{\sigma^3 c_{\text{eff}}^2}{\alpha} k_B T, \quad S^M_{\text{shot}}(\omega) \coloneqq \frac{\pi}{4} \frac{\sigma^2 \epsilon \tau_D}{l_D} k_B T,
\end{align}
where the effective capacitance is $\frac{1}{c_{\text{eff}}} = \frac{1}{c_0} + \frac{2l_D}{\epsilon}$. The term $S^M_{\text{JN}}(\omega)$ corresponds to Johnson-Nyquist noise, which is the thermal noise obtained in a continuum model of charge dynamics \cite{bryant_physical_2023}. We can identify the second term with shot noise, which depends only on the local fluctuations of the charge density, and is independent of the capacitance of the system. We use this expression to analyze $S^M_1(\omega)$ since the presence of the membrane does not significantly change the magnitude of this contribution:
\begin{align}
    S^M_{\text{shot}}(\omega) \le S^M_1(\omega) \le \frac{5}{2} S^M_{\text{shot}}(\omega).
\end{align}
In contrast, the Johnson-Nyquist term depends strongly on the capacitance of the membrane. For the parameters in the main text we have $c_0 \approx 10^{-2} \epsilon/l_D$, meaning that the effective capacitance is dominated by the membrane capacitance, so the Johnson-Nyquist fluctuations are heavily suppressed. We can evaluate when we would expect the shot noise contribution to dominate by evaluating at what value of $\sigma$ we have $S^{M}_{\text{JN}}(\omega) = S^M_{\text{shot}}(\omega)$. This value is given by
\begin{align}
    \frac{\sigma^*(d)}{l_D} = \frac{2}{\pi^{1/2}} \left( 1 + \frac{\epsilon d}{2 \epsilon_m l_D} \right)^2
\end{align}
Therefore, for $\sigma > \sigma^*(d)$, the system is the Johnson-Nyquist regime, and a continuum description of the charge dynamics becomes appropriate.

\begin{figure}
    \centering
    \begin{subfigure}[t]{0.5\textwidth}
        \centering
        \includegraphics[height=0.8\textwidth]{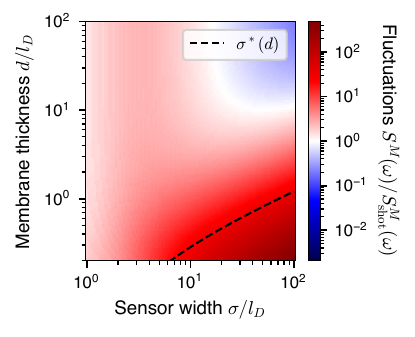}
        \caption{Comparison of fluctuations to shot noise.}
    \end{subfigure}
    ~ 
    \begin{subfigure}[t]{0.5\textwidth}
        \centering
        \includegraphics[height=0.8\linewidth]{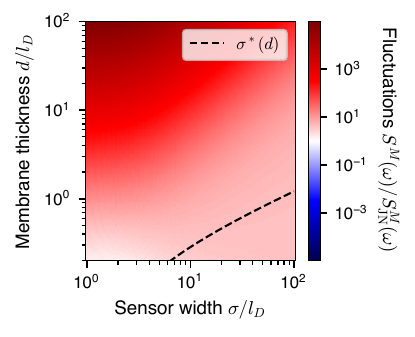}
        \caption{Comparison of fluctuations to Johnson-Nyquist noise.}
    \end{subfigure}
    \caption{Scaling of fluctuations for $\epsilon/\epsilon_m = 20$. The dashed line represents the values of $(\sigma,d)$ at which shot noise and Johnson-Nyquist noise coincide.}
    \label{fig:noise_param_sweep}
\end{figure}

\section{Sensing with a charged plate}
\subsection{Setup}
We now consider an alternate sensing mechanism in which the electric field, rather than the charge, is sensed. We consider a sensor composed of two plates, one with charge $Q$ and another of charge $-Q$. They both have a width $\sigma$ and are embedded inside the membrane. The displacement of the positive plate in the $z$ direction is $\xi$, and we impose that the negative plate has a displacement $-\xi$. We assume that there plate is confined mechanically by a quadratic potential, in addition to its interaction with the ionic charges. Therefore, the free energy of the system in the presence of the plate is
\begin{align}
    {\cal F} = {\cal F}_0 + \frac{K}{2} \xi^2 + \frac{Q}{2\pi \sigma^2} \int d^2{\bf x} \, e^{-\lVert {\bf x} \rVert^2/2\sigma^2} [\Phi^{\text{ions}}_m({\bf x}, z=\xi) - \Phi^{\text{ions}}_m({\bf x}, z=-\xi)],
\end{align}
where ${\cal F}_0$ is the free energy in Eq. \eqref{eq:original_free_energy} and $\Phi^{\text{ions}}$ is the potential generated by the ions.

We assume that the plate is overdamped and has mobility $\mu$, such that its dynamics are dictated by
\begin{align}
    \partial_t \xi = - \mu \left( \frac{\partial {\cal F}}{\partial \xi} \right).
\end{align}
With this specification, we can study the equilibrium fluctuations in $\xi$, which we take to be the sensor's measurement. These fluctuations determine the ability of the sensor to discriminate changes in voltage.

\subsection{Electric potential}
The movement of the plate causes movement in the ionic charge near the membrane, which then feeds back into its dynamics. Therefore, to begin our study of the dynamics of the plate, we need to specify how the presence of the plate affects the distribution of ions.

Let us consider a setup with a fixed ionic charge distribution $\rho$ and plate position $\xi$. The electric potential can be obtained as the sum of the potentials generated by the ionic charges and the plates, each satisfying the boundary conditions in Eqs. \eqref{eq:interface_conds} and \eqref{eq:potential_continuity}. Thus, we can use our solution in Eqs. \eqref{eq:bulk_potential} and \eqref{eq:membrane_potential} for $\Phi^{\text{ions}}$.

We can first solve for the potential generated by the positive plate. For this, we need to solve Poisson's equation:
\begin{align}
    (\partial_z^2 - k^2) \Phi^{\text{plate},+}({\bf k},z) = - \frac{Q}{\epsilon_m} e^{-k^2 \sigma^2/2} \delta(z-\xi).
\end{align}
In the bulk, we have exponentially-decaying solutions:
\begin{align}
    \Phi^{\text{plate}}_\pm({\bf k},z) = A_\pm e^{-(|z|-d/2)}.
\end{align}
Inside the plate, we can use the following bases, which satisfy the boundary conditions on the top and bottom interface, respectively,
\begin{align}
    L_+(z) &= \cosh(k(z-d/2)) - \frac{\epsilon}{\epsilon_m} \sinh(k(z-d/2)), \nonumber \\
    L_-(z) &= \cosh(k(z+d/2)) + \frac{\epsilon}{\epsilon_m} \sinh(k(z+d/2)).
\end{align}
Our solution inside the membrane, then, can be written as
\begin{align}
    \Phi^{\text{plate}}_m({\bf k}, z) =
    \begin{cases}
        A_+ L_+(z) & \xi < z < d/2, \\  
        A_- L_-(z) & -d/2 < z < \xi.
    \end{cases}
\end{align}
At the location of the plate ($z = \xi$), we have
\begin{align}
    \partial_z \Phi^{\text{plate},+}_m({\bf k}, \xi^+) - \partial_z \Phi^{\text{plate},+}_m({\bf k}, \xi^-) = - \frac{Q}{\epsilon_m} e^{-k^2 \sigma^2/2}.
\end{align}
Using this condition and continuity, we obtain the solution for the potential generated by the plate:
\begin{align}
    \Phi^{\text{plate},+}_m({\bf k}, z) = \frac{Q e^{-k^2 \sigma^2/2}}{\epsilon_m k \Delta}
    \begin{cases}
        L_-(\xi) L_+(z) & \xi < z < d/2, \\  
        L_+(\xi) L_-(z) & -d/2 < z < \xi,
    \end{cases}
\end{align}
where
\begin{align}
    \Delta \coloneqq \left( 1 + \frac{\epsilon^2}{\epsilon_m^2} \right) \sinh(kd) + \frac{2 \epsilon}{\epsilon_m} \cosh(kd).
\end{align}
To obtain the total potential generated by the sensor, we can use the symmetry of the problem and add both potentials:
\begin{align}
    \Phi^{\text{plate}}({\bf k}, z) = \Phi^{\text{plate},+}({\bf k}, z) + \Phi^{\text{plate},-}({\bf k}, z) = \Phi^{\text{plate},+}({\bf k}, z) - \Phi^{\text{plate},+}({\bf k}, -z).
\end{align}

\subsection{Charge dynamics}
In this section, we take $z$ to specify coordinates in the bulk for the sake of simplicity. Thus, $z = 0^\pm$ now denotes the bulk right above and below the membrane.

Since no charges are being introduced into the bulk, the charge dynamics we derived for our PNP analysis still hold, meaning that the evolution of charge is dictated by
\begin{align}
    \tau_D \partial_t \rho = -(1 + k^2 l_D^2) \rho + l_D^2 \partial_z^2 \rho.
\end{align}
Our no-flux condition still reads
\begin{align}
    \partial_z \rho({\bf k}, z=0^\pm) = \frac{\alpha}{D} E_z({\bf k}, z=0^{\pm}).
\end{align}

What changes in the presence of the plate is that the electric field now has a contribution from the plates. We will eventually be interested in the linear response regime for $\xi$, so here we characterize how the charges respond to a small perturbation in $\xi$. The field generated by the plates on either side of the membrane is
\begin{align}
    E^{\text{plate}}_z({\bf k}, 0^\pm) = -\partial_z \Phi^{\text{plate}}_z({\bf k}, 0^\pm) = \frac{Q e^{-k^2 \sigma^2/2}}{\epsilon_m \Delta} [L_-(\xi) - L_+(\xi)].
\end{align}
For small $\xi$, this is
\begin{align}
    E^{\text{plate}}_z({\bf k}, 0^\pm) = \frac{Q e^{-k^2 \sigma^2/2}}{\epsilon_m} \left( \frac{k \xi}{C + (\epsilon/\epsilon_m) S} \right) + {\cal O}(\xi^2),
\end{align}
where $C \coloneqq \cosh(kd/2)$ and $S \coloneqq \sinh(kd/2)$. 

Given the form of the electric field and our charge boundary conditions, we see that changing $\xi$ does not cause a change in the even component of the field, only on the odd one. Thus, we can focus on $\rho_{\text{odd}}$ for our response analysis. Using our previous characterization of the electric field generated by the ionic charges, we can write our odd boundary condition as
\begin{align}
    \partial_z \rho_{\text{odd}}({\bf k}, 0^+) = - \frac{\psi}{l_D^2} \int_0^\infty dz \, e^{-kz} \rho_{\text{odd}}({\bf k},z) + \frac{\alpha}{D} \frac{Q e^{-k^2 \sigma^2/2}}{\epsilon_m[C + (\epsilon/\epsilon_m) S]} k \xi.
\end{align}
Now, in the frequency domain, our dynamics take the form
\begin{align}
    l_D^2 \partial_z^2 \rho = (1 + k^2 l_D^2 + i \omega \tau_D) \rho \eqqcolon q^2 \rho.
\end{align}
Thus, our charge distribution is given by
\begin{align}
    \rho({\bf k}, z) = B e^{-qz/l_D}.
\end{align}
Using this ansatz in our boundary condition and solving for $B$, we obtain
\begin{align}
    B = \frac{\epsilon}{l_D} \left[ \frac{\psi}{q + kl_D} - q \right]^{-1} \frac{Q e^{-k^2 \sigma^2/2}}{\epsilon_m[C + (\epsilon/\epsilon_m) S]} k \xi.
\end{align}
This allows us to characterize the response of the charge distribution to changes in the plate position. We can now use this to study the full response of the system to perturbations.

\subsection{Response to perturbations}
To characterize the equilibrium fluctuations of $\xi$, we study the response of the system to a perturbation of the form
\begin{align}
    {\cal F} \to {\cal F} - r \xi.
\end{align}
The dynamics of the sensor under this perturbation are
\begin{align}
    \mu^{-1} \partial_t \xi = - K \xi - \frac{Q}{2 \pi \sigma^2} \int d^2 {\bf x} \, e^{-\lVert {\bf x} \rVert^2/2\sigma^2} [\partial_z \Phi^{\text{ions}}_m({\bf x}, \xi) + \partial_z \Phi^{\text{ions}}_m({\bf x}, -\xi)] + r.
\end{align}
Note that we can close these dynamics using our characterization of the electric field. For small perturbations $\xi$, we have
\begin{align}
    E_z^{\text{ions}}({\bf k},\xi) = - \frac{(\lambda^+({\bf k}) + \lambda^-({\bf k}))}{2 \epsilon_m C} - \frac{(\lambda^+({\bf k}) - \lambda^-({\bf k}))}{2 \epsilon_m S} k \xi + {\cal O}(\xi^2),
\end{align}
where $\lambda^\pm$ are defined in Eq. \eqref{eq:lambdas}. Thus, writing our dynamics in terms of a ${\bf k}$ integral, we obtain
\begin{align}
    \mu^{-1} \partial_t \xi = - K \xi - \frac{Q}{4 \pi^2 \epsilon_m} \int d^2 {\bf k} \, e^{-k^2 \sigma^2/2} \frac{(\lambda^+({\bf k}) + \lambda^-({\bf k}))}{C} + r.
\end{align}

The ionic interaction integral can be written in terms of the odd charge distribution:
\begin{align}
    \lambda^+({\bf k}) + \lambda^-({\bf k}) = 2 \psi \int_0^{\infty} dz \, \rho_{\text{odd}}({\bf k},z) e^{-kz}.
\end{align}
Using our characterization in the frequency domain, this is
\begin{align}
    \lambda^+({\bf k}) + \lambda^-({\bf k}) = \frac{2 \psi B}{k + q/l_D} = 2 \frac{\epsilon}{\epsilon_m} Q \left( \frac{\psi}{\psi - q (q + kl_D)} \right) \frac{e^{-k^2 \sigma^2/2} k \xi(\omega)}{C + (\epsilon/\epsilon_m) S}.
\end{align}
Define the effective electric spring constant as
\begin{align}
    K_{\text{el}} \coloneqq \frac{Q^2}{\pi \epsilon_m}  \frac{\epsilon}{\epsilon_m} \int_0^\infty dk \, e^{-k^2 \sigma^2} \frac{k^2}{C(C + (\epsilon/\epsilon_m) S)} \left( \frac{\psi}{\psi - q (q + kl_D)} \right).
\end{align}
With this, the dynamics of the sensor in frequency space are given by
\begin{align}
    i \omega \mu^{-1} \xi(\omega) = - (K + K_{\text{el}}) \xi(\omega) + r(\omega).
\end{align}
The response function to perturbations, then, is
\begin{align}
    \chi^{\xi r}(\omega) = \frac{\mu}{\mu(K + K_{\text{el}}) + i \omega}.
\end{align}

\subsection{The fluctuation spectrum}
Using the fluctuation-dissipation theorem, we have that the spectrum of $\xi$ fluctuations is
\begin{align}
    S^\xi(\omega) = - \frac{2 k_B T}{\omega} \text{Im}[\chi^{\xi r}(\omega)].
\end{align}
If $\omega \mu K, \omega \tau_D \ll 1$, then we can approximate the spectrum to leading order in $\omega$;
\begin{align}
    S^\xi(\omega) \approx \frac{2 k_B T}{\mu(K + K_{\text{el}})^2} + \frac{2 k_B T}{(K + K_{\text{el}})^2} \left. \frac{\partial K_{\text{el}}}{\partial (i \omega)} \right|_{\omega=0}.
\end{align}
To compute this last derivative, it is useful to write $K_{\text{el}}$ in terms of a dimensionless integral:
\begin{align}
    K_{\text{el}} &= \frac{Q^2}{\pi \epsilon_m l_D^3}  \frac{\epsilon}{\epsilon_m} \int_0^\infty d\kappa \, e^{-\kappa^2 (\sigma^2/l_D^2)} \frac{\kappa^2}{C(C + (\epsilon/\epsilon_m) S)} \left( \frac{\psi}{\psi - q (q + \kappa)} \right) \nonumber \\
    &\eqqcolon \frac{Q^2}{\pi \epsilon_m l_D^3}  \frac{\epsilon}{\epsilon_m} \int_0^\infty d\kappa \, e^{-\kappa^2 (\sigma^2/l_D^2)} {\cal I}(\kappa, q).
\end{align}
Since $\omega$ only enters $K_{\text{el}}$ through $q$, we can write our derivative as
\begin{align}
    \left. \frac{\partial K_{\text{el}}}{\partial (i \omega)} \right|_{\omega=0} = \frac{\tau_D}{2} \frac{Q^2}{\pi \epsilon_m l_D^3}  \frac{\epsilon}{\epsilon_m} \int_0^\infty d\kappa \, e^{-\kappa^2 (\sigma^2/l_D^2)} \frac{1}{q} \left. \frac{\partial {\cal I}}{\partial q} \right|_{q = (1+\kappa^2)^{1/2}}.
\end{align}

To make progress in analyzing our fluctuation spectrum, we can analyze the large $\sigma/l_D$ limit, as before. To do this, we expand our integrand around $\kappa = 0$. The first two leading terms are
\begin{align}
    \frac{1}{q} \left. \frac{\partial {\cal I}}{\partial q} \right|_{q = (1+\kappa^2)^{1/2}} = \frac{8}{\left( \frac{\epsilon}{\epsilon_m} \frac{d}{l_D} + 2 \right)^2} \left[ 1 - \frac{3}{2} \kappa \right] + {\cal O}(\kappa^2).
\end{align}
This give the following expansion of our fluctuation spectrum:
\begin{align}
    S^\xi(\omega) \approx \frac{2 k_B T}{\mu(K + K_{\text{el}})^2} + \frac{k_B T}{(K + K_{\text{el}})^2} \frac{4 \pi^{-1/2} Q^2 \tau_D}{l_D^3 \left( \frac{\epsilon}{\epsilon_m} \frac{d}{l_D} + 2 \right)^2} \frac{\epsilon}{\epsilon_m^2} \frac{l_D}{\sigma} \left[ 1 - \frac{3}{2 \pi^{1/2}} \frac{l_D}{\sigma} \right] + {\cal O}\left( \left( \frac{\sigma}{l_D} \right)^{-3} \right).
\end{align}
We can write this expression in terms of the effective capacitance of the membrane:
\begin{align}
    S^\xi(\omega) \approx \frac{2 k_B T}{\mu(K + K_{\text{el}})^2} + 4 \pi^{-1/2} \frac{k_B T}{(K + K_{\text{el}})^2} \frac{Q^2 \tau_D c_{\text{eff}}^2}{\sigma \epsilon_m^2 \epsilon} \left[ 1 - \frac{3}{2 \pi^{1/2}} \frac{l_D}{\sigma} \right] + {\cal O}\left( \left( \frac{\sigma}{l_D} \right)^{-3} \right).
\end{align}

We are interested in the fluctuations in a measurement of the voltage across the membrane. Assuming the mechanical spring dominates the effective ionic potential for low frequencies ($K \gg K_{\text{el}}(\omega \to 0)$), we can do a simple force balance to relate the measured potential to the spring displacement:
\begin{align}
    \frac{Q V}{d} \approx K \xi.
\end{align}
The fluctuations in the measurement of $V$, then, satisfy
\begin{align}
    S^V_{\text{plate}}(\omega) \approx \frac{K^2 d^2}{Q^2} S^\xi(\omega).
\end{align}
Using our result for the fluctuations in $\xi$ and using $K \gg K_{\text{el}}$, we obtain
\begin{align}
    S^V_{\text{plate}}(\omega) \approx \frac{2 d^2 k_B T}{\mu Q^2} + \frac{4 \pi^{-1/2}}{\alpha \sigma} \frac{c_{\text{eff}}^2}{c_0^2} k_B T \left[ 1 - \frac{3}{2 \pi^{1/2}} \frac{l_D}{\sigma} \right].
\end{align}
In the regime we are working in, we have $c_{\text{eff}} \approx c_0$, so our fluctuations are
\begin{align}
    S^V_{\text{plate}}(\omega) \approx \frac{2 d^2 k_B T}{\mu Q^2} + \frac{4 \pi^{-1/2}}{\alpha \sigma} k_B T \left[ 1 - \frac{3}{2 \pi^{1/2}} \frac{l_D}{\sigma} \right] \eqqcolon S^V_{\text{plate},0}(\omega) + S^V_{\text{plate},1}(\omega).
\end{align}

\subsection{Comparison to charge measurement}
In our bulk charge measurement calculation, the spectrum of fluctuations could be approximated by a Johnson-Nyquist noise and a shot noise contribution:
\begin{align}
    S^M_{\text{bulk}}(\omega) \approx S^M_{\text{JN}}(\omega) + S^M_{\text{shot}}(\omega).
\end{align}
We can translate these fluctuations to fluctuations in a measurement of the potential by looking at the estimate $V \approx M/2 \pi \sigma^2 c_{\text{eff}}$. Thus, fluctuations in voltage measurements inferred from the bulk have the following structure:
\begin{align}
    S^V_{\text{bulk}}(\omega) = \frac{\pi^{-1/2}}{\sigma \alpha} k_B T + \frac{1}{16 \pi} \frac{\epsilon \tau_D}{\sigma^2 l_D c_{\text{eff}}^2} k_B T.
\end{align}
Note that the Johnson-Nyquist contribution has the same scaling as the second term of $S^V_{\text{plate}}$. Intuitively, this is because Johnson-Nyquist noise comes from the stable modes used for communication. In contrast, the other sources of noise are qualitatively different in both schemes, arising from sensor implementation details. To compare these, it is useful to write the shot noise contribution in terms of the ion mobility  $\mu_{\text{ions}} = \beta D$. With this, the ratio between the fluctuation spectra is 
\begin{align}
    \frac{S^V_{\text{shot}}(\omega)}{S^V_{\text{plate},0}(\omega)} = \frac{1}{32 \pi} \frac{\epsilon^2}{\epsilon_m^2} \frac{Q^2}{\sigma^2 l_D \sum_i n_i^0 e^2 z_i^2} \frac{\mu}{\mu_{\text{ions}}}.
\end{align}
The relation between these fluctuations can be made more interpretable by defining the average square charge in the bulk sensor:
\begin{align}
    Q_{\text{bulk}}^2 \coloneqq \sigma^2 l_D \sum_i n_i^0 e^2 z_i^2.
\end{align}
Thus, we can write the ratio as
\begin{align}
    \frac{S^V_{\text{shot}}(\omega)}{S^V_{\text{plate},0}(\omega)} = \frac{1}{32 \pi} \frac{\epsilon^2}{\epsilon_m^2} \frac{Q^2}{Q^2_{\text{bulk}}} \frac{\mu}{\mu_{\text{ions}}}.
\end{align}
This characterization makes it clear that the first source of noise in the plate's measurements is of the same nature as our bulk shot noise. The bulk permittivity is around an order of magnitude higher than the membrane permittivity. Ion channels would need to implement an electric field sensor using charged residues, meaning that $Q$ is of the order of a few elementary charges and that the mobility $\mu$ cannot be much higher than the mobility of individual ions. This implies that this source of noise is a manifestation of shot noise in the plate's measurement arising from the interaction of its discrete components and the environment. 

\section{Collective sensing}
\subsection{Input signal and measurements}
We now study a continuum model of electrical communication on a spherical neuron. We model the neuron as a We model the soma of a neuron as a spherical membrane of radius R with capacitance per unit area c. As before, it is embedded in a bulk medium with conductivity $\alpha$. The charge density on the membrane is $\lambda({\bf x})$, where ${\bf x}$ now represent coordinates on the surface of the sphere, such that $\lVert {\bf x} \rVert = R$.

To model communication, we consider a ``sender'' located at the north pole of the sphere that injects a time-dependent electrical current $I(t)$ spread over an area of linear size $\sigma_I$. To capture this, we model the input current density $J({\bf x},t)$ using a von Mises-Fisher distribution:
\begin{align}
    J({\bf x},t) = I(t) \frac{e^{({\bf x} \cdot R \hat{{\bf z}})/\sigma_I^2}}{4 \pi \sigma_I^2 \sinh(R^2/\sigma_I^2)}.
\end{align}
The total current over the sphere is
\begin{align}
    \int ds({\bf x}) \, J({\bf x},t) = I(t),
\end{align}
where $ds({\bf x}) \coloneqq R^2 \sin(\theta) \, d\theta \, d\phi$ is the area element. In the $\sigma_I/R \to 0$ limit, where the system approximates an infinite plane, we recover a Gaussian spread of the current:
\begin{align}
    J({\bf x},t) \approx I(t) \frac{e^{({\bf x} \cdot R \hat{{\bf z}} - R^2) /\sigma_I^2}}{2 \pi \sigma_I^2} \approx I(t) \frac{e^{-\lVert x_\perp \rVert^2/2\sigma_I^2}}{2 \pi \sigma_I^2},
\end{align}
where ${\bf x}_\perp$ is the component of ${\bf x}$ perpendicular to $\hat{{\bf z}}$. 

Motivated by the von Mises-Fisher distribution, we define the input and measurement Gaussian distance factors between two vectors ${\bf x}$ and ${\bf y}$ as
\begin{align}
    \gamma^I({\bf x}, {\bf y}) \coloneqq e^{({\bf x} \cdot {\bf y} - R^2)/\sigma_I^2}, \quad \gamma^M({\bf x}, {\bf y}) \coloneqq e^{({\bf x} \cdot {\bf y} - R^2)/\sigma^2}.
\end{align}
The input current can be written in terms of this distance factor as
\begin{align} \label{eq:input_current}
    J({\bf x}, t) \approx I(t) \frac{\gamma^I({\bf x}, R\hat{{\bf z}})}{2 \pi \sigma_I^2}.
\end{align}
We consider two communication schemes. In the first, which we call the ``perfect instrument'', the charge over the entire cell is measured, so the output is $M_{\text{cell}}(t) = \int ds({\bf x}) \, \lambda({\bf x},t)$. The output of our second communication scheme is a vector of measurements $M_1(t), \ldots, M_N(t)$ made by $N$ ion channels at locations ${\bf y}_1, \ldots, {\bf y}_N$. The measurement of ion channel $j$ measures the charge in an area of linear size $\sigma$ around its position ${\bf y}_j$. In terms of our spherical distance, this is
\begin{align}
    M_j(t) = \int ds({\bf x}) \, \gamma^M({\bf x}, {\bf y}_j) \lambda({\bf x}).
\end{align}
Throughout the rest of the analysis, we assume that $\sigma, \sigma_I \ll R$.

\subsection{Continuum charge dynamics}
We now derive the dynamics in our continuum model of charge dynamics. We begin by analyzing the dynamics in the absence of noise. 

The potential difference between the inside and outside of the membrane is
\begin{align}
    \Phi_+({\bf x}, t) - \Phi_-({\bf x}, t) = \frac{\lambda({\bf x}, t)}{c},
\end{align}
where $\Phi_+$ and $\Phi_-$ represent the potential in the bulk outside and inside the membrane, respectively. The harmonic decomposition of $\Phi_+$ and $\Phi_-$ is
\begin{align}
    \Phi_\pm(r,\theta,\phi) = \sum_{\ell,m} \Phi^\pm_{\ell m}(r) Y_\ell^m(\theta,\phi).
\end{align}
In the bulk, the potential must satisfy Laplace's equation ($\nabla^2 \Phi = 0$). In spherical coordinates this is
\begin{align}
    \nabla^2 \Phi = \frac{1}{r^2} \frac{\partial}{\partial r} \left( r^2 \frac{\partial \Phi}{\partial r} \right) + \frac{1}{r^2 \sin(\theta)} \frac{\partial}{\partial \theta} \left( \sin(\theta) \frac{\partial \Phi}{\partial \theta} \right) + \frac{1}{r^2 \sin^2(\theta)} \frac{\partial^2 \Phi}{\partial \phi^2} = 0.
\end{align}
This equation has solutions for the harmonic components of the form
\begin{align} \label{eq:Phi_harmonics}
    \Phi^{\pm}_{\ell m}(r) = A^{\pm}_{\ell m} r^\ell + B^{\pm}_{\ell m} r^{-(\ell + 1)}.
\end{align}
Our first set of boundary conditions are that $\Phi$ must vanish as $r \to \infty$ and it must remain finite as $r \to 0$. This implies that $A_{\ell m}^+ = B_{\ell m}^- = 0$. To characterize the rest of the solution, we study how the bulk currents affect the charge density dynamics.

Bulk currents are given by Ohm's law: ${\bf J}_{\text{bulk}}({\bf x},t) = - \alpha \nabla \Phi$. At the membrane surface, charge accumulates due to bulk currents and drains from leak currents:
\begin{align}
    \partial_t \lambda({\bf x},t) = \alpha \left.\frac{\partial \Phi_+(r,{\bf x},t)}{\partial r} \right|_{r=R} + J_{\text{leak}}({\bf x},t) = \alpha \left.\frac{\partial \Phi_-(r,{\bf x},t)}{\partial r} \right|_{r=R} + J_{\text{leak}}({\bf x},t),
\end{align}
where the last equality is obtained by imposing that the charge inside the membrane has the same magnitude and opposite sign as the charge outside. Transforming this equation, we obtain an additional restriction on the coefficients:
\begin{align}
    \left. \frac{\partial \Phi^+_{\ell m}(r)}{\partial r} \right|_{r=R} = \left. \frac{\partial \Phi^-_{\ell m}(r)}{\partial r} \right|_{r=R}.
\end{align}
With this, the solution is fully specified. Using $J_{\text{leak}}({\bf x},t) = -\lambda({\bf x},t)/\tau_{\text{leak}}$, the dynamics of the charge density field are
\begin{align}
    \partial_t \lambda_\ell^m(t) = - \frac{1}{\tau_\ell} \lambda_\ell^m(t),
\end{align}
where the RC time of the $\ell$-th harmonic satisfies
\begin{align}
    \tau_\ell^{-1} = \tau_{\text{leak}}^{-1} + \frac{\alpha}{R c} \frac{\ell(\ell+1)}{2 \ell + 1}.
\end{align}
For the parameters on the main text, we have $\alpha \tau_{\text{leak}}/R c \approx 10^4$. Therefore, we have $\tau_\ell = \tau_{\text{leak}}$ for $\ell = 0$ and $\tau_\ell \approx \frac{Rc}{\alpha} \frac{2 \ell + 1}{\ell(\ell + 1)}$ for all $\ell > 0$. As we show next, this fast decay of the higher harmonics implies that communication is dominated by the $\ell = 0$ mode.

\subsection{Electrical signaling}
The effect of the sender's signal is to add an additional current to the dynamics of the charge density field. With the signaling current, the dynamics become
\begin{align}
    \partial_t \lambda_\ell^m(t) = - \frac{1}{\tau_\ell} \lambda_\ell^m(t) + J_\ell^m(t).
\end{align}
Writing these dynamics in frequency space and using the form of $J$ in Eq. \eqref{eq:input_current} yields
\begin{align}
    \lambda_\ell^m(\omega) = \frac{\gamma^I_{\ell m}(R \hat{{\bf z}}) \tau_\ell}{2 \pi \sigma_I^2 (1 + i \omega \tau_l)} I(\omega).
\end{align}
Here we have used $\gamma^I_{\ell m}({\bf y})$ to denote the harmonic components of $\gamma^I({\bf x}, {\bf y})$, transforming the first coordinate. Having the response of $\lambda_\ell^m$ to the input signal, we can characterize the signal transmitted in our two communication schemes.

\subsubsection{Whole-cell sensing}
In this case, our measurement is the total charge on the surface of the sphere. Note that we can write this as
\begin{align}
    M_{\text{cell}}(\omega) = \int ds({\bf x}) \, \lambda({\bf x},\omega) = 2 \pi^{1/2} R^2 \lambda_0^0(\omega) = \frac{R^2 \gamma^I_{00}(R \hat{{\bf z}}) \tau_{\text{leak}}}{\pi^{1/2} \sigma_I^2 (1 + i \omega \tau_{\text{leak}})} I(\omega).
\end{align}
Using Lemma \ref{lem:harmonic_distances} to approximate $\gamma^I_{00}(R \hat{{\bf z}})$, we obtain the response kernel of $M_{\text{cell}}(\omega)$ to $I(\omega)$:
\begin{align}
    \chi^{MI}_{\text{cell}}(\omega) \approx \frac{\tau_{\text{leak}}}{1 + i \omega \tau_{\text{leak}}}.
\end{align}

\subsubsection{Multiple sensors}
The measurement of channel $j$ in frequency space is
\begin{align}
    M_j(\omega) = \int ds({\bf x}) \, \gamma^M({\bf x}, {\bf y}_j) \lambda({\bf x}, \omega).
\end{align}
We can decompose the response of $M_j$ to $I$ into contributions form different harmonics. To do this, note that the measurement can be seen as a convolution of $\lambda$ and $\gamma^M$. The components of this convolution are computed in Lemma \ref{lem:spherical_convolution}, giving the following decomposition:
\begin{align}
    M_j(\omega) = \sum_{\ell,m} R^2 \frac{2 \pi^{1/2}}{(2 \ell + 1)^{1/2}} \gamma^M_{\ell 0}(R \hat{{\bf z}}) \lambda_\ell^m Y_\ell^m({\bf y}_j).
\end{align}
Now, note that $\lambda$ only has non-vanishing $m=0$ harmonics, since the signal is rotationally symmetric around $\hat{{\bf z}}$. Using the response of $\lambda_\ell^0$ to $I$ and the harmonic representation of distances in Lemma \ref{lem:harmonic_distances}, we get the following decomposition:
\begin{align}
    \chi^{MI}_j(\omega) \approx \frac{\sigma^2}{2R^2} \left( \frac{\tau_{\text{leak}}}{1 + i \omega \tau_{\text{leak}}} \right) + \frac{\sigma^2}{2R^2} \sum_{\ell > 0} \left( \frac{\tau_\ell}{1 + i \omega \tau_\ell} \right) (2 \ell + 1) e^{-(\ell+1/2)^2 (\sigma^2 + \sigma_I^2)/2 R^2} P_\ell(\cos(\theta_j)),
\end{align}
where $\chi^{MI}_j(\omega)$ is the response kernel of $M_j(\omega)$ to $I(\omega)$. 

For the parameters in the main text, we have $\tau_\ell^{-1} > \alpha/Rc \approx 10^{7}$ s${-1}$. Thus, for $\omega \ll 10^{7}$ s${-1}$, we have $\frac{\tau_\ell}{1 + i \omega \tau_\ell} \approx \tau_\ell \approx \frac{Rc}{\alpha} \frac{2 \ell + 1}{\ell(\ell+1)}$, which gives us the approximation
\begin{align}
    \chi^{MI}_j(\omega) \approx \frac{\sigma^2}{2R^2} \left( \frac{\tau_{\text{leak}}}{1 + i \omega \tau_{\text{leak}}} \right) + \frac{\sigma^2 c}{2 \alpha R} \sum_{\ell > 0} \frac{(2 \ell + 1)^2}{\ell(\ell+1)} e^{-(\ell+1/2)^2 (\sigma^2 + \sigma_I^2)/2 R^2} P_\ell(\cos(\theta_j)).
\end{align}
We can analyze the behavior of this response kernel for small angles. In this regime, the following approximation for the Legendre polynomials holds:
\begin{align} \label{eq:legendre_approximation}
    P_\ell(\cos (\theta)) = \sqrt{\frac{\theta}{\sin (\theta)}} J_0((\ell+1/2)\theta) + {\cal O}(\ell^{-1}),
\end{align}
where $J_0$ is the Bessel function of the first kind. Despite this being a large $\ell$ expansion, taking the leading order yields a very good approximation for small $\theta$, even for $\ell = 1$. Using this approximation and approximating the sum over $\ell$ with an integral, we obtain
\begin{align}
    \chi^{MI}_j(\omega) \approx \frac{\sigma^2}{2R^2} \left( \frac{\tau_{\text{leak}}}{1 + i \omega \tau_{\text{leak}}} \right) + \frac{2 \sigma^2 c}{\alpha R} \sqrt{\frac{\theta_j}{\sin (\theta_j)}} \int_{0}^{\infty} du \, e^{-u^2 (\sigma^2 + \sigma_I^2)/2 R^2} J_0(u \theta_j).
\end{align}
We can now use the identity (Eq. 6.631.1 in \cite{gradshtein_table_2007})
\begin{align} \label{eq:bessel_integral}
    \int_0^\infty du \, e^{-au^2} J_0(bx) = \frac{1}{2} \left( \frac{\pi}{a} \right)^{1/2} e^{-b^2/8a} I_0 \left( \frac{b^2}{8a} \right),
\end{align}
where $I_0$ is the modified Bessel function of the first kind. Thus, our approximation now is
\begin{align}
    \chi^{MI}_j(\omega) \approx \frac{\sigma^2}{2R^2} \left( \frac{\tau_{\text{leak}}}{1 + i \omega \tau_{\text{leak}}} \right) + \frac{\pi^{1/2} \sigma c}{\alpha} \left( \frac{2 \sigma^2}{\sigma^2 + \sigma_I^2} \right)^{1/2} \sqrt{\frac{\theta_j}{\sin (\theta_j)}} e^{-R^2 \theta_j^2/4(\sigma^2 + \sigma_I^2)} I_0\left( \frac{R^2 \theta_j^2}{4(\sigma^2 + \sigma_I^2)} \right).
\end{align}
The function $e^{-x} I_0(x)$ behaves as $(2 \pi x)^{-1/2}$ for large $x$. Therefore, for $\frac{\max\{\sigma, \sigma_I\}}{R} \ll \theta \ll 1$, the expression above scales as
\begin{align}
    \chi^{MI}_j(\omega) \approx \frac{\sigma^2}{2R^2} \left( \frac{\tau_{\text{leak}}}{1 + i \omega \tau_{\text{leak}}} \right) + \frac{2 \sigma c}{\alpha} \frac{\sigma}{R \theta_j}
\end{align}
If $\omega \tau_{\text{leak}}$ is ${\cal O}(1)$, then for $\theta_j > \frac{Rc}{\alpha \tau_{\text{leak}}} \approx 10^{-4}$ the second term in the response function is dominated with respect to the first. Thus, the signal for most of the cell is driven by the zeroth harmonic:
\begin{align}
    \chi^{MI}_j(\omega) \approx \frac{\sigma^2}{2R^2} \left( \frac{\tau_{\text{leak}}}{1 + i \omega \tau_{\text{leak}}} \right) = \frac{\sigma^2}{2R^2} \chi^{MI}_{\text{cell}}(\omega).
\end{align}

\subsection{Johnson-Nyquist noise}
We now calculate the fluctuations in our continuum model of charge dynamics. To do this, we use the fluctuation-dissipation theorem. The energy of the system is the energy stored in the membrane capacitor. We perturb this energy by coupling the charge density $\lambda$ to a conjugate field $h$ in the Hamiltonian:
\begin{align}
    {\cal H}[\lambda, h] = \int ds({\bf x}) \, \left[ \frac{1}{2} \frac{\lambda^2({\bf x})}{c} - h({\bf x}) \lambda({\bf x}) \right].
\end{align}
We do this so we can characterize the response of the charge to the perturbation, which can be linked to the noise spectrum via the fluctuation-dissipation theorem. The functional derivative of the perturbed Hamiltonian is
\begin{align}
    \frac{\delta {\cal H}[\lambda, h]}{\delta \lambda({\bf x})} = \frac{\lambda({\bf x})}{c} - h({\bf x}).
\end{align}
The perturbed dynamics are obtained by making the substitution
\begin{align}
    \frac{\lambda({\bf x})}{c} \to \frac{\delta {\cal H}[\lambda, h]}{\delta \lambda({\bf x})}.
\end{align}
Finally, our perturbed dynamics (in the absence of a signal) are
\begin{align} \label{eq:perturbed_dynamics}
    \partial_t \lambda_\ell^m(t) = - \frac{1}{\tau_\ell} \lambda_\ell^m(t) - \frac{c}{\tau_\ell} h_\ell^m(t).
\end{align}

To apply the fluctuation-dissipation theorem, we need to compute the response function of $\lambda$ to $h$. We can do this by writing Eq. \eqref{eq:perturbed_dynamics} in Fourier space, which yields the response function
\begin{align}
    \chi^{\lambda h}_{\ell m}(\omega) = -\frac{c}{1 + i \omega \tau_\ell}.
\end{align}
The fluctuation-dissipation theorem applied to the spherical setting is given in Lemma \ref{lem:spherical_FDT}. Using our response function, the fluctuations in $\lambda$ are
\begin{align} \label{eq:lambda_fluctuation_components}
    \tilde{S}^\lambda_{\ell m}(\omega) = (2 \ell + 1)^{1/2} \frac{c \tau_\ell}{\pi^{1/2} R^2 (1 + \omega^2 \tau_\ell^2)} \delta_{m0} k_B T.
\end{align}
Explicitly, $\tilde{S}^\lambda({\bf x}, \omega)$ is the Fourier transform of $\Tilde{C}^\lambda({\bf x}, t) \coloneqq C^\lambda({\bf x} , R \hat{{\bf z}}, t)$, where
\begin{align}
    C^\lambda({\bf x}, {\bf y}, t) \coloneqq \langle \lambda({\bf x}, t_0) \lambda({\bf y}, t_0 + t) \rangle.
\end{align}
This expectation is taken with respect to the equilibrium fluctuations of the system. With the structure of the fluctuations of $\lambda$, we cna compute the fluctuations in our measurements.

\subsubsection{Whole-cell sensing}
First, let us compute the temporal correlation function of the measurements of the cell
\begin{align}
    C^M_{\text{cell}}(t) = \langle M_{\text{cell}}(t_0) M_{\text{cell}}(t_0 + t) \rangle.
\end{align}
Writing out the measurement of the cell explicitly,
\begin{align}
    C^M_{\text{cell}}(t) = \int ds({\bf x}) \, ds({\bf y}) \, \langle \lambda({\bf x}, t_0) \lambda({\bf y}, t_0+t) \rangle = \int ds({\bf x}) \, ds({\bf y}) \, C^\lambda({\bf x}, {\bf y}, t).
\end{align}
Since the correlation function can only depend on ${\bf x} \cdot {\bf y}$, we can use rotational symmetry to reduce the expression above to a single integral:
\begin{align}
    C^M_{\text{cell}}(t) = 4 \pi R^2 \int ds({\bf x}) \, C^\lambda({\bf x}, R \hat{{\bf z}}, t) = 4 \pi R^2 \int ds({\bf x}) \, \tilde{C}^\lambda({\bf x}, t) = 8 \pi^{3/2} R^4 \tilde{C}^\lambda_{00}(t).
\end{align}
Taking the Fourier transform and using Eq. \eqref{eq:lambda_fluctuation_components}, the fluctuation spectrum of the cell measurement is
\begin{align}
    S^M_{\text{cell}}(\omega) = \frac{8 \pi R^2 c \tau_{\text{leak}}}{(1 + \omega^2 \tau_{\text{leak}}^2)} k_B T.
\end{align}

\subsubsection{Multiple sensors}
First, we compute the fluctuations in the measurement made by a single sensor. Since there is no signal, we can set ${\bf y}_j = R\hat{{\bf z}}$ due to rotational symmetry. The correlations in the channel measurement are
\begin{align}
    C^M_{\text{channel}}(t) &= \int ds({\bf x}) \, ds({\bf y}) \, \gamma^M({\bf x}, R\hat{{\bf z}}) \gamma^M({\bf y}, R\hat{{\bf z}}) \langle \lambda({\bf x}, t_0) \lambda({\bf y}, t_0+t) \rangle \nonumber \\
    &= \int ds({\bf x}) \, ds({\bf y}) \, \gamma^M({\bf x}, R\hat{{\bf z}}) \gamma^M({\bf y}, R\hat{{\bf z}}) C^\lambda({\bf x}, {\bf y}, t)
\end{align}
We can write the ${\bf y}$ integral as a convolution and use Lemma \ref{lem:spherical_convolution} to represent its components. This gives us a decomposition of the correlation function:
\begin{align}
    C^M_{\text{channel}}(t) = 2 \pi^{1/2} R^4 \sum_{\ell, m} \frac{(-1)^m}{(2 \ell + 1)^{1/2}} \gamma^M_{\ell m}(R \hat{{\bf z}}) \gamma^M_{\ell, -m}(R \hat{{\bf z}}) \tilde{C}^\lambda_{\ell 0}(t).
\end{align}
Taking the Fourier transform and using Lemma \ref{lem:harmonic_distances} to write our distance factors, we get a decomposition of the channel noise spectrum:
\begin{align}
    S^M_{\text{channel}}(\omega) \approx 2 \pi^{3/2} \sigma^4 \left[ \tilde{S}^\lambda_{00}(\omega) + \sum_{\ell > 0} (2 \ell + 1)^{1/2} e^{-(\ell+1/2)^2\sigma^2/R^2} \tilde{S}^\lambda_{\ell 0}(\omega) \right].
\end{align}
With our expression for the $\lambda$ spectrum, this decomposition becomes
\begin{align}
    S^M_{\text{channel}}(\omega) \approx \frac{2 \pi c \sigma^4}{R^2} k_B T \left[ \frac{\tau_{\text{leak}}}{1 + \omega^2 \tau_{\text{leak}}^2} + \sum_{\ell > 0} \frac{(2 \ell + 1) \tau_\ell}{1 + \omega^2 \tau_\ell^2} e^{-(\ell+1/2)^2\sigma^2/R^2} \right].
\end{align}

The decomposition above is useful to understand the sources of noise in the channel's measurements. The first term corresponds to the noise from the measurement $(\sigma^2/2R^2) M_{\text{cell}}$, so we can interpret it as the collective noise coming from the fluctuations in the zeroth mode. The rest of the decomposition corresponds to the local Johnson-Nyquist fluctuations. To see this, suppose that $\omega$ is small compared to the first few $\tau_\ell$. Then we can approximate the sum as
\begin{align}
    \sum_{\ell > 0} \frac{(2 \ell + 1) \tau_\ell}{1 + \omega^2 \tau_\ell^2} e^{-(\ell+1/2)^2\sigma^2/R^2} \approx \frac{Rc}{\alpha} \sum_{\ell > 0} (2 \ell + 1) \frac{2 \ell + 1}{\ell(\ell + 1)} e^{-(\ell+1/2)^2\sigma^2/R^2}.
\end{align}
We can further approximate this by evaluating the sum as an integral:
\begin{align} \label{eq:integral_approx}
    \sum_{\ell > 0} \frac{(2 \ell + 1) \tau_\ell}{1 + \omega^2 \tau_\ell^2} e^{-(\ell+1/2)^2\sigma^2/R^2} \approx \frac{Rc}{\alpha} \int_1^{\infty} du \, \frac{(2 u + 1)^2}{u(u + 1)} e^{-(u+1/2)^2\sigma^2/R^2} \approx \frac{R^2 c}{\alpha \sigma} \int_0^{\infty} dv \frac{(2v)^2}{v^2} e^{-v^2}.
\end{align}
Evaluating this integral and substituting into our expression of fluctuations, we get
\begin{align}
    S^M_{\text{channel}}(\omega) \approx \frac{2 \pi c \sigma^4}{R^2} \left( \frac{\tau_{\text{leak}}}{1 + \omega^2 \tau_{\text{leak}}^2} \right) k_B T + \frac{4 \pi^{3/2} c^2 \sigma^3}{\alpha } k_B T.
\end{align}
The second term is precisely the single-channel Johnson-Nyquist from Equation \eqref{eq:noise_scales}. Therefore, a single ion channel experiences noise from the slow fluctuations in the zeroth harmonic and from the local fluctuations we characterized in Section \ref{sec:measurement_fluctuations}.

To conclude our characterization of noise in the multiple-channel measurement, we show that fluctuations in different channels are only correlated through the fluctuations in the total charge. For this, we compute the correlation function between the measurements of channels $j$ and $k$:
\begin{align}
    C^M_{jk}(t) = \langle M_j(t_0) M_k(t_0 + t) \rangle.
\end{align}
Due to rotational symmetry, we can perform a rotation such that ${\bf y}_j$ gets sent to $R \hat{{\bf z}}$. Under this rotation, ${\bf y}_k$ gets sent to a new position, which we denote ${\bf y}_{jk}$. Following the same procedure as for our previous noise spectra, we find the following decomposition of the cross spectrum:
\begin{align}
    S^M_{jk}(\omega) = 2 \pi^{1/2} R^4 \sum_\ell (2 \ell + 1)^{-1/2} \gamma^M_{\ell 0}(R \hat{{\bf z}}) \gamma^M_{\ell 0}({\bf y}_{jk}) \tilde{S}^\lambda_{\ell 0}(\omega).
\end{align}
Using Lemma \ref{lem:arbitrary_harmonics}, we can obtain the scaling of this spectrum in $\theta_{jk}$, the angular distance between ${\bf y}_j$ and ${\bf y}_k$:
\begin{align}
    S^M_{jk}(\omega) \approx \frac{2 \pi c \sigma^4}{R^2} \left( \frac{\tau_{\text{leak}}}{1 + \omega^2 \tau_{\text{leak}}^2} \right) + \frac{2 \pi c \sigma^4}{R^2} k_B T \sum_{\ell > 0} \frac{\tau_\ell}{1 + \omega^2 \tau_\ell^2} (2 \ell + 1) e^{-(\ell+1/2)^2 \sigma^2/R^2} J_0((\ell + 1/2) \theta_{jk}).
\end{align}
Note that for small $\omega$ this sum is very similar to the one in Equation \eqref{eq:integral_approx}, with an added Bessel function. Therefore, we can perform the same integral approximation and use Eq. \eqref{eq:bessel_integral} to obtain
\begin{align}
    \sum_{\ell > 0} \frac{(2 \ell + 1) \tau_\ell}{1 + \omega^2 \tau_\ell^2} e^{-(\ell+1/2)^2\sigma^2/R^2} J_0((\ell + 1/2) \theta_{jk}) &\approx \frac{4 R^2 c}{\alpha \sigma} \int_0^{\infty} dv \, e^{-v^2} J_0\left( \frac{R}{\sigma} v \theta_{jk} \right) \nonumber \\
    &= \frac{2 \pi^{1/2} R^2 c}{\alpha \sigma} e^{-\theta_{jk}^2 R^2/8 \sigma^2} I_0\left( \frac{\theta_{jk}^2 R^2}{8 \sigma^2} \right).
\end{align}
Thus, our cross-spectrum has the following behavior:
\begin{align}
    S^M_{jk}(\omega) \approx \frac{2 \pi c \sigma^4}{R^2} \left( \frac{\tau_{\text{leak}}}{1 + \omega^2 \tau_{\text{leak}}^2} \right) k_B T + \frac{4 \pi^{3/2} c^2 \sigma^3}{\alpha } k_B T \cdot e^{-\theta_{jk}^2 R^2/8 \sigma^2} I_0\left( \frac{\theta_{jk}^2 R^2}{8 \sigma^2} \right).
\end{align}
The function $e^{-x} I_0(x)$ behaves as $(2 \pi x)^{-1/2}$ for large $x$, so the local Johnson-Nyquist contribution is suppressed if the channels that are separated more than a few $\sigma$. Therefore, for separated channels only the first term survives. This cross-spectrum is consistent with the model in the main text, in which every channel experiences a common source of noise from the zeroth harmonic and independent local noise. Since we know shot noise dominates the local component, we can assume the local measurement noise is simply shot noise.
\section{Technical Lemmas}

\subsection{Harmonic representation of distances} \label{lem:harmonic_distances}
\begin{tcolorbox}[colback=blue!10, title=Lemma: $\gamma^{I,M}_{\ell m}(R \hat{{\bf z}})$ formula]
The gaussian distance factor $\gamma^I({\bf x}, R\hat{{\bf z}})$ has harmonic components
\begin{align}
    \gamma^I_{\ell m}(R \hat{{\bf z}}) = \delta_{m 0} \pi^{1/2} (2 \ell + 1)^{1/2} \left[ \int_{-1}^1 du \, e^{-(1-u)R^2/\sigma_I^2} P_\ell(u) \right],
\end{align}
where $P_\ell$ is the $\ell$-th Legendre polynomial. Additionally, in the $\sigma_I/R \ll 1$ regime, these can be approximated by
\begin{align}
    \gamma^I_{\ell 0}(R \hat{{\bf z}}) \approx
    \begin{cases}
        \pi^{1/2} (\sigma_I^2/R^2) & \textrm{if } \ell = 0, \\
        \pi^{1/2} (\sigma_I^2/R^2) (2 \ell + 1)^{1/2} e^{-(\ell + 1/2)^2 \sigma_I^2/2 R^2} & \textrm{if } \ell > 0.
    \end{cases}
\end{align}
Analogous results hold for $\gamma^M$, replacing $\sigma_I$ with $\sigma$.
\end{tcolorbox}

The harmonic component $\gamma^I_{\ell m}(R \hat{{\bf z}})$ is given by
\begin{align}
    \gamma^I_{\ell m}(R \hat{{\bf z}}) = \frac{1}{R^2} \int ds({\bf x}) \, \gamma^I({\bf x}, R \hat{{\bf z}}) Y_\ell^{m*}({\bf x}).
\end{align}
Since $\gamma^I({\bf x}, R \hat{{\bf z}})$ has rotational symmetry about the north pole, we have that all the $m \ne 0$ components vanish. For the $m = 0$ components, we can write the integral in terms of Legendre polynomials using spherical coordinates:
\begin{align}
    \gamma^I_{\ell m}(R \hat{{\bf z}}) &= \frac{e^{-R^2/\sigma_I^2}}{R^2}  \delta_{m 0} \int ds({\bf x}) \, e^{R {\bf x} \cdot \hat{{\bf z}}/\sigma_I^2} Y_\ell^{0*}({\bf x}) \nonumber \\
    &= \frac{e^{-R^2/\sigma_I^2}}{R^2} \delta_{m 0} (2 \ell + 1)^{1/2} \int_0^\pi d \theta \, \sin(\theta) e^{R^2 \cos(\theta)/\sigma_I^2} P_{\ell}(\cos(\theta)).
\end{align}
Substituting $u = \cos(\theta)$, we obtain the desired result:
\begin{align}
    \gamma^I_{\ell m}(R \hat{{\bf z}}) = \delta_{m 0} \pi^{1/2} (2 \ell + 1)^{1/2} \left[ \int_{-1}^1 du \, e^{-(1-u)R^2/\sigma_I^2} P_\ell(u) \right].
\end{align}

For $\ell = 0$, we have $P_0(u) = 1$. Since we are working in the $\sigma_I/R \ll 1$ regime, we can make the following approximation:
\begin{align}
    \int_{-1}^1 du \, e^{-(1-u)R^2/\sigma_I^2} \approx \int_{0}^\infty dv \, e^{-v R^2/\sigma_I^2} = \frac{\sigma_I^2}{R^2}.
\end{align}
Therefore, our zeroth harmonic can be approximated by
\begin{align}
    \gamma^I_{\ell 0}(R \hat{{\bf z}}) = \pi^{1/2} \frac{\sigma_I^2}{R^2}.
\end{align}

For higher harmonics, we can approximate $\gamma^I_{\ell 0}$ by using the asymptotic behavior of the Legendre polynomials:
\begin{align}
    P_\ell(\cos (\theta)) = \sqrt{\frac{\theta}{\sin (\theta)}} J_0((\ell+1/2)\theta) + {\cal O}(\ell^{-1}),
\end{align}
where $J_0$ is the Bessel function of the first kind. Despite this being a large $\ell$ expansion, taking the leading order yields a very good approximation for small $\theta$, even for $\ell = 1$. Thus,
\begin{align}
    \int_{-1}^1 du \, e^{-(1-u) R^2/\sigma_I^2} P_\ell(u) \approx \int_0^\pi d\theta \, \sqrt{\theta \sin(\theta)} \, e^{-(1-\cos(\theta)) R^2/\sigma_I^2} J_0((\ell+1/2) \theta).
\end{align}
Again, since $\sigma_I/R \ll 1$, most of the mass concentrates near $\theta = 0$, meaning that we can take the upper limit to $\infty$ and expand the functions that do not depend on $\ell$ to leading order:
\begin{align}
    \int_{-1}^1 du \, e^{-(1-u) R^2/\sigma_I^2} P_\ell(u) \approx \int_0^\infty dy \, y e^{-y^2 R^2/2\sigma_I^2} J_0((\ell+1/2) y) = \frac{\sigma_I^2}{R^2} e^{-(\ell+1/2)^2 \sigma_I^2/2 R^2}.
\end{align}
This yields our approximation for $\ell > 0$:
\begin{align}
    \gamma^I_{\ell 0}(R\hat{{\bf z}}) \approx \pi^{1/2} \frac{\sigma_I^2}{R^2} (2 \ell + 1)^{1/2} e^{-(\ell+1/2)^2 \sigma_I^2/2 R^2}.
\end{align}

\subsection{Spherical Convolution} \label{lem:spherical_convolution}
\begin{tcolorbox}[colback=blue!10, title=Lemma: Spherical Convolution]
Let $f$ be defined by
\begin{align}
    f({\bf x}) = \int ds({\bf x}') \, h({\bf x}') g({\bf x} \cdot {\bf x}').
\end{align}
Then the harmonic components of $f$ are
\begin{align}
    f_\ell^m = R^2 \sqrt{\frac{4 \pi}{2 \ell + 1}} h_\ell^m \tilde{g}_\ell^0,
\end{align}
where $\Tilde{g}({\bf x}) \coloneqq g({\bf x} \cdot R \hat{{\bf z}})$.
\end{tcolorbox}

To simplify the proof, we will use Dirac notation. For a given function $f({\bf x})$, define the associated state $\ket{f}$ by
\begin{align}
    \ket{f} \equiv \sum_{\ell,m} f_\ell^m \ket{\ell,m}.
\end{align}
Additionally, we have that the space representation of this state is given by $f({\bf x}) = \braket{{\bf x}|f}$, where the states $\ket{{\bf x}}$ satisfy the completeness relation
\begin{align}
    \frac{1}{R^2} \int ds({\bf x}) \, \ket{{\bf x}}\bra{{\bf x}} = \mathbbm{1}.
\end{align}
We can, therefore, write the integral expression for $f$ as
\begin{align}
    \braket{{\bf x}|f} = \int ds({\bf x}') \, \braket{{\bf x}'|h} g({\bf x} \cdot {\bf x}').
\end{align}
By making a change of variables ${\bf x}' \to {\bf A}({\bf x}) {\bf x}'$ using a rotation matrix ${\bf A}({\bf x})$ that satisfies ${\bf A}^{-1}({\bf x}) {\bf x} = R \hat{{\bf z}}$, we can write the integral as
\begin{align}
    \braket{{\bf x}|f} &= \int ds({\bf x}') \, \braket{{\bf A}({\bf x}) {\bf x}'|h} g(R {\bf x}' \cdot \hat{{\bf z}}) \nonumber \\
    &= \int ds({\bf x}') \, \braket{{\bf A}({\bf x}) {\bf x}'|h} \tilde{g}({\bf x}') \nonumber \\
    &= \int ds({\bf x}') \, \braket{{\bf A}({\bf x}) {\bf x}'|h} \braket{{\bf x}'|\Tilde{g}}.
\end{align}
Since $h$ is a real function, it must satisfy $\braket{{\bf A}({\bf x}) {\bf x}'|h} = \braket{h | {\bf A}({\bf x}) {\bf x}'}$. Additionally, we have that in state space, rotations act on kets as $\ket{{\bf A}({\bf x}) {\bf x}'} = {\cal A}({\bf x}) \ket{{\bf x}'}$, where ${\cal A}({\bf x})$ is the associated rotation operator. Using this and the completeness relation, we obtain
\begin{align}
    \braket{{\bf x} | f} = R^2 \braket{h | {\cal A}({\bf x}) | \tilde{g}}.
\end{align}
In order to get the harmonic components of $f$, we integrate the above expression, which yields
\begin{align}
    \braket{\ell,m|f} &= \int ds({\bf x}) \, \braket{\ell,m|{\bf x}} \braket{h|{\cal A}({\bf x})|\tilde{g}} \nonumber \\
    &= \sum_{\ell',m'} \sum_{\ell'',m''} \braket{h|\ell',m'} \braket{\ell'',m''| \tilde{g}} \int ds({\bf x}) \, \braket{\ell,m|{\bf x}} \braket{\ell',m'|{\cal A}({\bf x})|\ell'',m''}.
\end{align}
Now, note that, since $\Tilde{g}$ has rotational symmetry about the north pole, $\braket{\ell'',m''|\Tilde{g}} = \braket{\ell'',0|\Tilde{g}} \delta_{m'' 0}$. This implies that
\begin{align}
    \braket{\ell,m|f} = \sum_{\ell',m'} \sum_{\ell''} \braket{h|\ell',m'} \braket{\ell'',0|\Tilde{g}} \int ds({\bf x}) \, \braket{\ell,m|{\bf x}} \braket{\ell',m'|{\cal A}({\bf x})|\ell'',0}.
\end{align}
Since the subspace associated to the $\ell$ harmonics is invariant under rotations, we know that $\braket{\ell',m'|{\cal A}({\bf x})|\ell'',0} \propto \delta_{\ell'\ell''}$. we, then, have that
\begin{align}
    \braket{\ell,m|f} = \sum_{\ell',m'} \braket{h|\ell',m'} \braket{\ell',0|g} \int ds({\bf x}) \, \braket{\ell,m|{\bf x}} \braket{\ell',m'|{\cal A}({\bf x})|\ell',0}.
\end{align}

Let $\ket{R \hat{{\bf z}}}$ denote the position state corresponding to the north pole. We have that, for a given $\ell$, the only spherical harmonic that has a non-zero component at the north pole corresponds to $m=0$. Specifically, we have that
\begin{align}
    \braket{R \hat{{\bf z}}|\ell,m} = \sqrt{\frac{2 \ell + 1}{4 \pi}} \delta_{m0} = \sqrt{\frac{2 \ell + 1}{4 \pi}} \braket{\ell,0|\ell,m}.
\end{align}
Additionally, note that since ${\bf x} = {\bf A}({\bf x}) (R \hat{{\bf z}})$, we can write
\begin{align}
    \braket{{\bf x} | \ell,m} = \braket{R \hat{{\bf z}}|{\cal A}({\bf x})^\dag|\ell,m}.
\end{align}
Since the $\ell$ subspace is invariant under rotations, we have that ${\cal A}({\bf x})^\dag\ket{\ell,m}$ is a linear combination of $\ell$ harmonics. Given the previous result, this implies that
\begin{align}
    \braket{{\bf x} | \ell,m} = \sqrt{\frac{2 \ell + 1}{4 \pi}} \braket{\ell,0|{\cal A}({\bf x})^\dag|\ell,m}.
\end{align}
Taking the conjugate of this expression, we obtain
\begin{align}
    \braket{\ell,m|{\bf x}} = \sqrt{\frac{2 \ell + 1}{4 \pi}} \braket{\ell,m|{\cal A}({\bf x})|\ell,0}.
\end{align}
Plugging this into the expression for $f_\ell^m$:
\begin{align}
    \braket{\ell,m|f} &= \sum_{\ell',m'} \sqrt{\frac{4 \pi}{2 \ell' + 1}} \braket{h|\ell',m'} \braket{\ell',0|\tilde{g}} \int ds({\bf x}) \, \braket{\ell,m|{\bf x}} \braket{\ell',m'|{\bf x}}.
\end{align}
The spherical harmonics satisfy
\begin{align}
    \braket{\ell,m|{\bf x}} = (-1)^m \braket{{\bf x}|\ell,-m}.
\end{align}
This yields the following expression:
\begin{align}
    \braket{\ell,m|f} &= \sum_{\ell',m'} (-1)^{m'} \sqrt{\frac{4 \pi}{2 \ell' + 1}} \braket{h|\ell',m'} \braket{\ell',0|\tilde{g}} \int ds({\bf x}) \, \braket{\ell,m|\theta,\phi} \braket{\theta,\phi|\ell',-m'} \nonumber \\
    &= R^2 \sum_{\ell',m'} (-1)^{m'} \sqrt{\frac{4 \pi}{2 \ell' + 1}} \braket{h|\ell',m'} \braket{\ell',0|\tilde{g}} \braket{\ell,m|\ell',-m'} \nonumber \\
    &= R^2 (-1)^{m} \sqrt{\frac{4 \pi}{2 \ell + 1}} \braket{h|\ell,-m} \braket{\ell,0|\tilde{g}} \nonumber \\
    &= R^2 \sqrt{\frac{4 \pi}{2 \ell + 1}} \braket{\ell,m|h} \braket{\ell,0|\tilde{g}},
\end{align}
where the last equality follows from the same conjugation property used before. Switching back to the original notation, we have that
\begin{align}
    f_\ell^m = R^2 \sqrt{\frac{4 \pi}{2 \ell + 1}} h_\ell^m \Tilde{g}_\ell^0.
\end{align}
Note that $f$ corresponds to the isotropic convolution of $h$ and $g$, as defined in \cite{kennedy_azimuthally_2011}.

\subsection{Spherical fluctuation-dissipation theorem} \label{lem:spherical_FDT}
\begin{tcolorbox}[colback=blue!10, title=Lemma: Fluctuation-dissipation theorem on a sphere]
Let a system have a hamiltonian given by the following functional:
\begin{align}
    H[\lambda] = H_0[\lambda] - \int ds({\bf x}) \, \lambda({\bf x}) h({\bf x}),
\end{align}
where $h$ is the conjugate field. Additionally, suppose that the harmonic components have the following impulse-response behavior:
\begin{align} \label{eq:harmonic_response}
    \langle \lambda_\ell^m(t) \rangle = \langle \lambda_\ell^m \rangle_0 + \int_{-\infty}^t d\tau \, h_\ell^m(\tau) \chi^{\lambda h}_{\ell m}(t-\tau),
\end{align}
where $\langle \cdot \rangle_0$ denotes the expectation in the absence of a field. Then the spectrum of the equilibrium fluctuations in the $\lambda$ field is given by
\begin{align}\label{eq:fluctuation_dissipation}
    \tilde{S}^\lambda_{\ell m}(\omega) = - \sqrt{\frac{2 \ell + 1}{4 \pi}} \frac{2 k_B T}{\omega R^2} \mathrm{Im}\left[\chi^{\lambda h}_{\ell 0}(\omega) \right] \delta_{m0},
\end{align}
where $\tilde{S}^\lambda({\bf x}, \omega)$ is the Fourier transform of $\Tilde{C}^\lambda({\bf x}, t) \coloneqq C^\lambda({\bf x} , R \hat{{\bf z}}, t)$, and $C^\lambda$ is the equilibrium correlation function.
\end{tcolorbox}

Consider the case where the $h$ field is active for all $t \le 0$ and is turned off at $t = 0$:
\begin{align}
    h({\bf x},t) = h_0({\bf x}) \Theta(-t),
\end{align}
where $\Theta$ is the Heaviside step function. For $t > 0$, we can write the expectation of the $\lambda$ field as
\begin{align}
    \langle \lambda({\bf x},t) \rangle = \int {\cal D} \lambda' \, {\cal D} \lambda \, \lambda'({\bf x}) {\cal P}(\lambda',t|\lambda,0) {\cal W}(\lambda,0),
\end{align}
where ${\cal P}$ is the transition kernel and ${\cal W}$ is given by the Boltzmann distribution at $t=0$:
\begin{align}
    {\cal W}(\lambda,0) = \frac{e^{-\beta H[\lambda]}}{\int {\cal D} \lambda' \, e^{-\beta H[\lambda']}}.
\end{align}
For a weak field, we have that
\begin{align}
    {\cal W}(\lambda,0) \approx {\cal W}_0(\lambda,0) \left[1 + \beta \int ds({\bf x}) \, (\lambda({\bf x}) - \langle \lambda({\bf x}) \rangle_0) h_0({\bf x}) \right].
\end{align}
Returning to the expression for $\langle \lambda \rangle$,
\begin{align}
    \langle \lambda({\bf x},t) \rangle &= \int {\cal D} \lambda' \, {\cal D} \lambda \, \lambda'({\bf x}) {\cal P}(\lambda',t|\lambda,0) {\cal W}_0(\lambda,0) \left[1 + \beta \int ds({\bf x}') \, (\lambda({\bf x}') - \langle \lambda({\bf x}') \rangle_0) h_0({\bf x}') \right] \nonumber \\
    &= \langle \lambda({\bf x}) \rangle_0 + \beta \int ds({\bf x}') \, h_0({\bf x}') [\langle \lambda({\bf x}',0) \lambda({\bf x},t) \rangle_0 - \langle \lambda({\bf x}) \rangle_0 \langle \lambda({\bf x}') \rangle_0].
\end{align}
We can define the correlation function in equilibrium as
\begin{align}
    C^\lambda({\bf x}, {\bf x}',t,t') = \langle \lambda({\bf x},t) \lambda({\bf x}',t') \rangle_0 - \langle \lambda({\bf x}) \rangle_0 \langle \lambda({\bf x}') \rangle_0.
\end{align}
Note that this correlation function should be invariant under rotations and under time translations, meaning that it can be fully specified by ${\bf x} \cdot {\bf x}'$, and by $|t-t'|$. We, therefore, define
\begin{align}
    C^\lambda({\bf x}\cdot{\bf x}',|t-t'|) = \langle \lambda({\bf x},t) \lambda({\bf x}',t') \rangle_0 - \langle \lambda({\bf x}) \rangle_0 \langle \lambda({\bf x}') \rangle_0.
\end{align}
Plugging this into our previous expression yields
\begin{align}
    \langle \lambda({\bf x},t) \rangle = \langle \lambda({\bf x}) \rangle_0 + \beta \int ds({\bf x}') \, h_0({\bf x}') C^\lambda({\bf x} \cdot {\bf x}',t).
\end{align}
Using Lemma \eqref{lem:spherical_convolution}, we can write the integral as
\begin{align}
    \int ds({\bf x}') \, h_0({\bf x}') C^\lambda({\bf x} \cdot {\bf x}', t) &= R^2 \sum_{\ell, m} \sqrt{\frac{4 \pi}{2 \ell + 1}} (h_0)_{\ell}^m \tilde{C}^\lambda_{\ell 0}(t) Y_\ell^m({\bf x}),
\end{align}
where $\Tilde{C}^\lambda({\bf x}, t) = C^\lambda(R {\bf x} \cdot \hat{{\bf z}}, t)$. We, therefore, obtain the following expression for the harmonic components of $\lambda$:
\begin{align}
    \langle \lambda_\ell^m(t) \rangle = \langle \lambda_\ell^m \rangle_0 + \beta R^2 \sqrt{\frac{4 \pi}{2 \ell + 1}} (h_0)_\ell^m \tilde{C}^\lambda_{\ell 0}(t).
\end{align}
Comparison with Equation \eqref{eq:harmonic_response} yields
\begin{align}
    \int_{-\infty}^0 d\tau \, (h_0)_\ell^m \chi^{\lambda h}_{\ell m}(t-\tau) = \beta R^2 \sqrt{\frac{4 \pi}{2 \ell + 1}} (h_0)_\ell^m \tilde{C}^\lambda_{\ell 0}(t).
\end{align}
For this to be valid for any conjugate field, we must have that
\begin{align}
    \int_{-\infty}^0 d\tau \, \chi^{\lambda h}_{\ell m}(t-\tau) = \beta R^2 \sqrt{\frac{4 \pi}{2 \ell + 1}} \tilde{C}^\lambda_{\ell 0}(t).
\end{align}

Since the correlation function is even in $t$ and real, the power spectrum must satisfy
\begin{align}
    \Tilde{S}^\lambda_{\ell 0}(\omega) = 2 \mathrm{Re}\left[ \int_0^{\infty} dt \, e^{-i\omega t} \tilde{C}^\lambda_{\ell 0}(t) \right].
\end{align}
Using the previous expression for the correlation function, we have that
\begin{align}
    \tilde{S}^\lambda_{\ell 0}(\omega) &= \sqrt{\frac{2 \ell + 1}{4 \pi}} \frac{2 k_B T}{R^2} \mathrm{Re} \left[ \int_0^{\infty} dt \, e^{-i\omega t} \int_{0}^\infty d\tau \, \chi^{\lambda h}_{\ell m}(t+\tau) \right] \nonumber \\
    &= \sqrt{\frac{2 \ell + 1}{4 \pi}} \frac{2 k_B T}{R^2} \mathrm{Re} \left[ \int_0^{\infty} dt \, e^{-i\omega t} \int_{0}^\infty d\tau \, \chi^{\lambda h}_{\ell m}(\tau) \Theta(\tau-t) \right] \nonumber \\
    &= \sqrt{\frac{2 \ell + 1}{4 \pi}} \frac{2 k_B T}{R^2} \mathrm{Re} \left[ \frac{i}{\omega} \int_{0}^\infty d\tau \, \chi^{\lambda h}_{\ell m}(\tau) \left( e^{-i \omega \tau} - 1 \right) \right] \nonumber \\
    &= \sqrt{\frac{2 \ell + 1}{4 \pi}} \frac{2 k_B T}{\omega R^2} \mathrm{Re} \left[ i \left( \chi^{\lambda h}_{\ell m}(\omega) - \chi^{\lambda h}_{\ell m}(0) \right) \right].
\end{align}
Since $\chi^{\lambda h}_{\ell m}(0)$ is real, we have that
\begin{align}
    \tilde{S}^\lambda_{\ell 0}(\omega) = - \sqrt{\frac{2 \ell + 1}{4 \pi}} \frac{2 k_B T}{\omega R^2} \mathrm{Im}\left[\chi^{\lambda h}_{\ell m}(\omega) \right].
\end{align}
Since this is valid for any $m$, we evaluate it specifically for $m=0$.

\subsection{Harmonic representation of distances for arbitrary orientations} \label{lem:arbitrary_harmonics}
\begin{tcolorbox}[colback=blue!10, title=Lemma: $\gamma^{I,M}_{\ell 0}({\bf y})$ formula]
The gaussian distance factor $\gamma^I({\bf x}, {\bf y})$ has $m=0$ harmonic components
\begin{align}
    \gamma^I_{\ell 0}({\bf y}) = \gamma^I_{\ell 0}(R \hat{{\bf z}}) P_\ell(\cos(\theta)),
\end{align}
where $P_\ell$ is the $\ell$-th Legendre polynomial and $\theta$ is the polar angle of ${\bf y}$. For $\sigma_I/R \ll 1$ and $\theta \ll 1$, these can be approximated by
\begin{align}
    \gamma^I_{\ell 0}({\bf y}) \approx
    \begin{cases}
        \pi^{1/2} (\sigma_I^2/R^2) & \textrm{if } \ell = 0, \\
        \pi^{1/2} (\sigma_I^2/R^2) (2 \ell + 1)^{1/2} e^{-(\ell + 1/2)^2 \sigma_I^2/2 R^2} J_0((\ell + 1/2) \theta) & \textrm{if } \ell > 0.
    \end{cases}
\end{align}
Analogous results hold for $\gamma^M$, replacing $\sigma_I$ with $\sigma$.
\end{tcolorbox}

Since $\gamma^I({\bf x},{\bf y})$ only depends on ${\bf x} \cdot {\bf y}$, we can write
\begin{align}
    \gamma^I({\bf x},{\bf y}) \eqqcolon g\left(\frac{{\bf x}}{R} \cdot \frac{{\bf y}}{R} \right).
\end{align}
That is, $g$ is a function of $\cos(\theta)$, where $\theta$ is the angular distance between ${\bf x}$ and ${\bf y}$. Since the Legendre polynomials are complete on $[-1,1]$, $g$ can be expanded as
\begin{align}
    g\left(\frac{{\bf x}}{R} \cdot \frac{{\bf y}}{R} \right) = \sum_{\ell} a_\ell P_\ell\left(\frac{{\bf x}}{R} \cdot \frac{{\bf y}}{R} \right).
\end{align}
The Legendre polynomials satisfy the following identity:
\begin{align}
    P_\ell \left(\frac{{\bf x}}{R} \cdot \frac{{\bf y}}{R} \right) = \frac{4 \pi}{2 \ell + 1} \sum_{m=-\ell}^\ell Y_\ell^m({\bf x}) Y_\ell^{m*}({\bf y}).
\end{align}
Therefore, $\gamma^I({\bf x},{\bf y})$ admits the decomposition
\begin{align}
    \gamma^I({\bf x},{\bf y}) = \sum_{\ell,m} \frac{4 \pi}{2 \ell + 1} a_\ell Y_\ell^m({\bf x}) Y_\ell^{m*}({\bf y}).
\end{align}
Taking this expression as a function of ${\bf x}$, its harmonic components are
\begin{align}
    \gamma^I_{\ell m}({\bf y}) = \frac{4 \pi}{2 \ell + 1} a_\ell Y_\ell^{m*}({\bf y}).
\end{align}

To obtain the coefficients $a_\ell$, we can evaluate our expression at ${\bf y} = R \hat{{\bf z}}$:
\begin{align}
    \gamma^I_{\ell m}(R \hat{{\bf z}}) = \frac{4 \pi}{2 \ell + 1} a_\ell Y_\ell^{m*}(R \hat{{\bf z}}) = \frac{2 \pi^{1/2}}{(2 \ell + 1)^{1/2}} a_\ell \delta_{m 0} \implies a_\ell = \frac{(2 \ell + 1)^{1/2}}{2 \pi^{1/2}} \gamma^I_{\ell 0}(R \hat{{\bf z}}).
\end{align}
Using this expression in our harmonic components for $\gamma^I$ and evaluating at $m = 0$, we obtain
\begin{align}
    \gamma^I_{\ell 0}({\bf y}) = \frac{2 \pi^{1/2}}{(2 \ell + 1)^{1/2}} \gamma^I_{\ell 0}(R \hat{{\bf z}}) Y_\ell^{0*}({\bf y}) = \gamma^I_{\ell 0}(R \hat{{\bf z}}) P_\ell(\cos(\theta)).
\end{align}

Our approximation is obtained by approximating $\gamma^I_{\ell 0}(R \hat{{\bf z}})$ using Lemma \ref{lem:harmonic_distances} and approximating the Legendre polynomials by
\begin{align}
    P_\ell(\cos(\theta)) \approx \sqrt{\frac{\theta}{\sin(\theta)}} J_0((\ell + 1/2) \theta).
\end{align}
For small angles, we have
\begin{align}
    \sqrt{\frac{\theta}{\sin(\theta)}} = 1 + {\cal O}(\theta^2),
\end{align}
so we can use the approximation $P_\ell(\cos(\theta)) \approx J_0((\ell + 1/2) \theta)$.

\end{document}